\documentclass[aps,prd,amssymb,amsmath,amsfonts,nofootinbib,eqsecnum,reprint,showpacs,longbibliography,groupedaddress]{revtex4-1}

\usepackage{graphicx}
\usepackage{lmodern}
\usepackage{amsmath,amssymb,amsfonts}
\usepackage{mathrsfs}
\usepackage[utf8]{inputenc}
\usepackage[T1]{fontenc}
\usepackage{url}
\usepackage[colorlinks]{hyperref}
\usepackage[dvipsnames]{xcolor}
\usepackage{multirow}
\usepackage[normalem]{ulem}
\usepackage{orcidlink}
\usepackage{placeins}
\usepackage{mathtools}
\usepackage{delimset}
\usepackage{empheq}
\usepackage[caption=false]{subfig}
\usepackage{placeins}
\usepackage{cleveref}
\usepackage{float}

\Crefname{table}{Tab.}{Tabs.}

\graphicspath{ {./figures/}}

\definecolor{dodgerblue}{HTML}{1E90FF}
\definecolor{viennared}{HTML}{DA0A14}
\definecolor{ctorange}{HTML}{FF6C0C}
\definecolor{wales}{HTML}{ff0038}
\definecolor{benettongreen}{HTML}{009421}
\definecolor{ferrarired}{HTML}{ff2800}
\definecolor{austriawienpurple}{HTML}{441678}
\definecolor{steiermarkgruen}{HTML}{006747}

\hypersetup{
     colorlinks=true,
     linkcolor=dodgerblue,
     filecolor=dodgerblue,
     citecolor = dodgerblue,      
     urlcolor=dodgerblue,
     }

\newcommand{\br}{\bar{r}}
\newcommand{\brmin}{\bar{r}_{\rm min}}
\newcommand{\pinf}{p_{\infty}}

\newcommand{\bu}{\bar{u}}

\newcommand{\bfp}{\boldsymbol{p}}
\newcommand{\bfr}{\boldsymbol{r}}

\newcommand{\Birmingham}{School of Physics and Astronomy and Institute for Gravitational Wave Astronomy, University of Birmingham, Edgbaston, Birmingham, B15 2TT, United Kingdom}

\begin{document}

\title{Strong Field Scattering of Black Holes: Assessing Resummation Strategies}

\author{Shaun Swain \orcidlink{0009-0001-8487-0358}} 
\email{sxs2159@student.bham.ac.uk}
\affiliation{\Birmingham}

\author{Geraint Pratten \orcidlink{0000-0003-4984-0775}} 
\email{g.pratten@bham.ac.uk}
\affiliation{\Birmingham}

\author{Patricia Schmidt \orcidlink{0000-0003-1542-1791}} 
\email{p.schmidt@bham.ac.uk}
\affiliation{\Birmingham}

\date{\today}

\begin{abstract}
Recent developments in post-Minkowskian (PM) calculations have led to a fast-growing body of weak-field perturbative information. As such, there is major interest within the gravitational wave community as to how this information can be used to improve the accuracy of theoretical waveform models. In this work, we build on recent efforts to validate high-order PM calculations using numerical relativity simulations. 
We present a new set of high-energy scattering simulations for equal-mass, non-spinning binary black holes, further expanding the existing suite of numerical relativity (NR) simulations.
We outline the basic features of three recently proposed resummation schemes (the $\mathcal{L}$-resummed model, the $w^\mathrm{eob}$ model and the SEOB-PM model) and compare the analytical predictions to our NR data. 
All of the models struggle to accurately capture the behavior at high energies, with common features including PM hierarchical shifts and divergences. 
The NR data is used to calibrate pseudo-5PM corrections to the scattering angle and to inform the EOB radial potentials. 
In each case, we argue that including higher-order information improves the agreement between the analytical models and NR, though the extent of improvement depends on how this information is incorporated and the choice of analytical baseline. 
Finally, we demonstrate that further resummation of the EOB radial potentials could be an effective strategy to improving the model agreement.

\end{abstract}

\maketitle 

\section{Introduction}
\label{sec:introduction}

Further advancements in gravitational wave (GW) detector technology \cite{Reitze:2019iox, Punturo:2010zz, LISA:2017pwj} will enable high precision measurements of GW signals, such as those performed by the LVK collaboration \cite{LIGOScientific:2018mvr,LIGOScientific:2020ibl,KAGRA:2021vkt,Nitz:2021zwj,Olsen:2022pin}. To decipher information from these signals, further theoretical developments are required to push the computational efficiency and accuracy of current waveform models beyond their capabilities. Successful characterisation of these events enable inference of astrophysical properties, as well as allowing strong-field tests of gravity.

Numerical relativity (NR) simulations of compact binaries \cite{Pretorius:2005gq, Campanelli:2005dd, Baker:2005vv, Jani:2016wkt, Boyle:2019kee, Healy:2020vre, Ramos-Buades:2022lgf, Hamilton:2023qkv, Bonino:2024xrv} provide a non-perturbative, numerical prediction for the gravitational waveform. Whilst NR simulations have played a key role in the rapid development of gravitational wave astronomy, there exist numerous technical challenges. Accuracy and run-time are currently limited by several factors: accuracy of initial data, robustness of gauge choices, efficient numerical methods, and so forth. 
In addition, NR simulations are poorly suited to GW data analysis due to the run time per simulation and the discrete coverage of the parameter space. 
To address these challenges, NR surrogate models \cite{Field:2013cfa, Blackman:2017dfb, Varma:2019csw} have been created using a data-driven approach, whereby a basis is built directly from the NR waveforms. Upon coupling with interpolation methods, a waveform can be reconstructed at any point in the parameter space spanned by the NR data. 
However, surrogate models implicitly inherit restrictions due to the limited parameter space coverage and duration of the NR simulations.

One way to overcome these limitations is to use a hybrid approach that combines analytical approximations with the numerical simulations, enabling a blend of perturbative and non-perturbative information.
The most common analytical tool to date has been the post-Newtonian (PN) approximation \cite{Blanchet:2013haa}, which consists of a perturbative expansion of the Einstein field equations in the small velocity $[\epsilon \sim (v/c) \ll 1]$ and weak field $[\epsilon \sim GM / (rc^2) \ll 1]$ limits. 
Two families of models which have successfully built on the PN approximation are: i) effective one-body (EOB) models, e.g. \cite{Buonanno:1998gg,Buonanno:2000ef,Damour:2001tu,Nagar:2018zoe,Nagar:2019wds,Nagar:2020pcj,Pompili:2023tna,Khalil:2023kep}, which are constructed from an NR calibrated EOB Hamiltonian containing PN information, and ii) phenomenological models \cite{Ajith:2007kx,Ajith:2009bn,Pratten:2020ceb,Garcia-Quiros:2020qpx,Pratten:2020fqn}, which directly model the emitted GW signal by calibrating phenomenological coefficients against PN/NR or EOB/NR hybrid waveforms. 

An alternative scheme is provided by the post-Minkowskian (PM) approximation \cite{Blanchet:2013haa}, which is valid in the weak field regime $[\epsilon \sim GM / (rc^2) \ll 1]$ \textit{but} relaxes the low velocity restriction. 
Interest in the PM approximation was revitalized by the pioneering work of \cite{Damour:2016gwp,Damour:2017zjx}. In~\cite{Damour:2016gwp}, it was realized that the usual EOB map between the PN description of two-body bound states and the bound states of a test particle could be replaced by a map expressed in terms of \textit{classical} scattering states. 
In practice, this means that the classical scattering function can be used to inform the EOB Hamiltonian for two gravitationally interacting bodies. 
The map was further expanded in~\cite{Damour:2017zjx} to derive the 2PM Hamiltonian for point-masses, together with a route to connecting classical scattering calculations to the quantum gravitational scattering amplitude of two particles.
This triggered significant activity, as the EOB framework could now leverage all the machinery that had been developed to calculate quantum scattering amplitudes at high perturbative orders, e.g. building on generalized unitarity \cite{Bern:1994zx} or the perturbative duality between gravity and gauge theories \cite{Bern:2008qj,Bern:2010yg,Bern:2010ue,Monteiro:2014cda}.

The PM approximation is naturally suited to gravitational scattering at large separations, and 
many important physical quantities have now been computed in the PM framework using a range of theoretical techniques: scattering amplitudes \cite{Cheung:2018wkq, Guevara:2018wpp, Kosower:2018adc, Bern:2019nnu, Bern:2019crd, Bjerrum-Bohr:2019kec, Herrmann:2021tct,Bern:2021dqo,Bern:2021yeh,Bjerrum-Bohr:2021din,Manohar:2022dea,Saketh:2021sri}, worldline field theory \cite{Mogull:2020sak,Riva:2021vnj,Jakobsen:2021smu, Jakobsen:2022psy,Bini:2021gat,Bini:2022wrq,Bini:2022enm, Driesse:2024xad}, effective field theory \cite{Kalin:2020mvi, Kalin:2020fhe, Mougiakakos:2021ckm, Dlapa:2021npj, Dlapa:2021vgp, Kalin:2022hph, Dlapa:2022lmu} and eikonalization \cite{KoemansCollado:2019ggb, DiVecchia:2019kta, DiVecchia:2021bdo}. 
Another important aspect is that the $(n+1)$PM order expansion inherently includes all $n$PN order information, together with additional information at infinitely high PN orders corresponding to high velocities. This makes the PM expansion particularly advantageous for systems where the component velocities can become large, such as binaries with significant eccentricity \cite{Chiaramello:2020ehz,Ramos-Buades:2022lgf,Nagar:2021gss,Ramos-Buades:2021adz} or hyperbolic encounters \cite{Cho:2018upo,Albanesi:2024xus}. As such, there has been a dedicated effort in exploring how we can use the PM expansion to inform the properties of generic bound systems, e.g. \cite{Damour:2016gwp,Damour:2017zjx, Kalin:2019rwq, Kalin:2019inp, Saketh:2021sri, Khalil:2022ylj, Buonanno:2024byg}. 

It is imperative to validate the PM information prior to its adoption in waveform models. That is, to understand how many PM orders are necessary for accurate waveform predictions throughout the parameter space. One route to do so is to use NR simulations to compare analytical predictions against non-perturbative information extracted from the simulations.  
Taking steps in this direction, \cite{Damour:2014afa,Damour:2022ybd, Rettegno:2023ghr, Buonanno:2024vkx} have compared scattering angles from NR simulations of equal mass binary black hole scattering events (both spinning and non-spinning) to analytical predictions of the scattering angle using resummations of the current PM information.
Building on this work, we significantly expand the suite of NR scattering simulations with a particular focus on exploring the high-energy behaviour in the equal-mass, non-spinning limit. Using these simulations, we present a detailed comparison of several approaches to resumming analytical PM information, highlighting the benefits and challenges posed by each of the schemes. 

The outline of this paper is as follows: In Sec. \ref{Section: Numerical Relativity} we present new high energy NR simulations of equal mass, non-spinning binary black hole scattering events and outline a robust procedure for extraction of the scattering angles. In Sec. \ref{Section: Relativistic Scattering Angles and the EOB Framework}, we recall general properties of unbound encounters before reviewing scattering angles in the PM and EOB frameworks. Sec. \ref{Section: Comparisons to NR} documents our results, whereby we review several attempts to resum the scattering angle and compare the analytical predictions against NR data. Following this, we suggest potential avenues of improvement where applicable.

\subsection*{Notation} \label{notation}
Here, we outline a number of useful quantities and conventions used throughout the paper. 
Unless otherwise stated, we adopt geometric units in which $G = c = 1$. 
The notation has been chosen to align with that adopted in, for example, \cite{Damour:2022ybd,Rettegno:2023ghr}.
In this paper, we only consider non-spinning two-body systems with black hole masses $m_1$ and $m_2$, such that the mass ratio is given by $q = m_1 / m_2 \geq 1$. 
Several useful mass parameters naturally follow
\begin{align}
M \! &= m_1 + m_2, \!\! & \nu \! &= \! \frac{\mu}{M} = \frac{m_1 m_2}{M^2}, 
\end{align}
where $M$ is the total mass, $\nu$ the symmetric mass ratio, and $\mu$ the reduced mass. 
The total energy of the system will be denoted by $E = E_1 + E_2$ together with the corresponding dimensionless energy
\begin{align}
\Gamma &= \frac{E}{M} = \sqrt{ 1 + 2 \nu (\gamma - 1)}.
\end{align}
Here we have introduced the relativistic Lorentz factor
\begin{align}
\gamma = \frac{1}{\sqrt{1 - v^2}},
\end{align}
in terms of the relativistic relative velocity $v$ such that $\gamma > 1$. The Lorentz factor is related to the dimensionless effective energy $\hat{E}_{\rm eff} = E_{\rm eff} / \mu$ by
\begin{align}
\gamma &= \hat{E}_{\rm eff} = \frac{E^2 - m^2_1 - m^2_2}{2 m_1 m_2}. 
\end{align}

The dynamics of the binary can be described in terms of mass-rescaled coordinates and momenta
\begin{align}
r &\equiv R / (G M), &t &\equiv T / (G M), \\ 
p_{\alpha} &= P_{\alpha} / \mu , & j &\equiv J / (G \mu M),
\end{align}
where $P^{\alpha}$ is the 4-momentum, $R$ the separation, $T$ the time, and $J$ the total angular momentum. 
In the center-of-mass (c.m.) frame, the momentum at past infinity $p_{\infty}$ is defined by \cite{Damour:2017zjx}
\begin{align}
p_{\infty} &= \sqrt{\gamma^2 - 1},
\end{align}
noting that this differs by a factor of $\mu$ from other conventions, e.g \cite{Buonanno:2024byg,Buonanno:2024vkx}. The relative position and momentum vectors, $\bfr$ and $\bfp$, describe the two-body dynamics following
\begin{align}
\bfp^2 &= p_r^2 + \frac{\ell^2}{r^2},
\end{align}
where $p_r = \hat{\bfr} \cdot \bfp$ and 
\begin{align}
\ell = \frac{L}{G \mu M },
\end{align}
is the rescaled orbital angular momentum. 
Note that as $\ell^{-1} \sim \mathcal{O}(G)$, we can express the perturbative PM expansion in terms of powers of $G$ or, for example, in terms of inverse powers of the angular momentum. 
We define the critical angular momentum, $\ell_0$, as the value of $\ell$ at which the two body system transitions from unbound $(\ell > \ell_0)$ to bound $(\ell < \ell_0)$ orbits. Below $\ell_0$, the black holes will emit radiation and eventually plunge.

In the non-spinning limit, $L$ is equivalent to the total angular momentum $J$ and we will typically use the canonical definition of the orbital angular momentum, i.e. $L = | \bfr \times \textbf{P} |$, following the Newton-Wigner spin-supplementary condition \cite{Newton:1949wig}. 
However, it can also be convenient to work with the covariant angular-momentum $L_{\rm cov} = b | \textbf{P} |$, where $b = \sqrt{-b^{\mu} b_{\mu}}$ denotes the impact parameter, following the Tulczyjew-Dixon spin supplementary condition \cite{Tulczyjew:1959zza,Dixon:1970zza}. 
This is particularly true when considering the extension to spinning binaries.

\section{Numerical Relativity}
\label{Section: Numerical Relativity}

In \cite{Damour:2014afa, Rettegno:2023ghr}, NR simulations of non-spinning, equal mass binary black hole scattering events were presented at energies of $\Gamma_1 = 1.02264$, $\Gamma_2 = 1.04032$ and $\Gamma_3 = 1.05548$. We extend this suite of NR simulations by performing 4 new sets of high-energy simulations at the following energies
\begin{align*}
\Gamma_4 &= 1.07277, \\
\Gamma_5 &= 1.11346, \\
\Gamma_6 &= 1.16174, \\
\Gamma_7 &= 1.21688.
\end{align*}
The Lorentz factors and related quantities are detailed in Tab.~\ref{tab:NR_quantities}. The numerical setup adopted in this work is templated on that of \cite{Damour:2014afa, Rettegno:2023ghr}, where we will highlight changes to the previous setup as appropriate.

\begin{figure}
    \centering
    \includegraphics[width=1\linewidth]{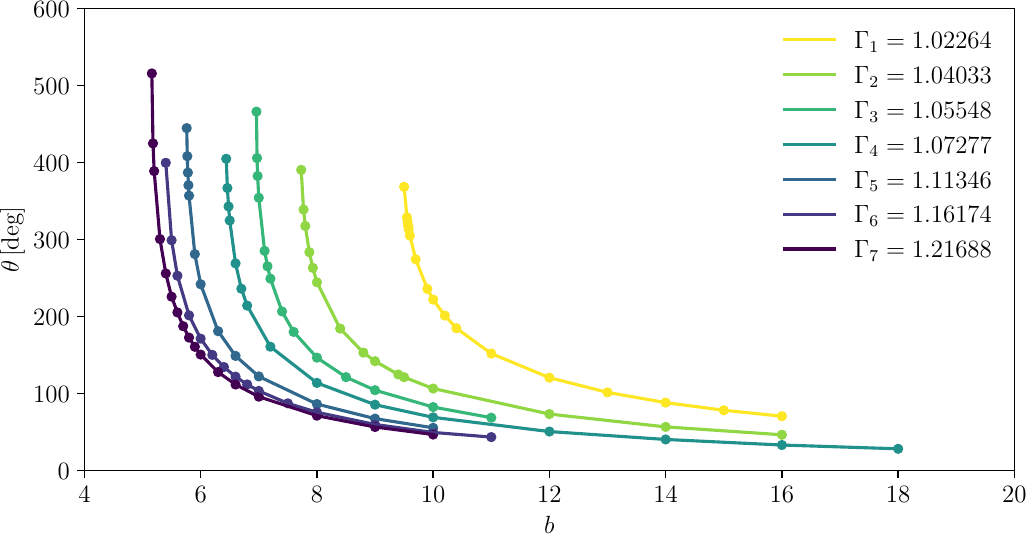}
    \caption{Up-to-date catalogue of scattering angles from non-spinning, equal mass binary black hole encounters as a function of the impact parameter. Data points at $\Gamma_1$ were gathered from \cite{Rettegno:2023ghr}, based on the earlier simulations presented in \cite{Damour:2014afa}, while those at $\Gamma_2$ and $\Gamma_3$ were provided by \cite{Rettegno:2023ghr}.
    }
    \label{fig:all_nr_data}
\end{figure}

\begin{table}[htbp]
    \caption{Summary of the energies considered for the current catalogue of non-spinning, equal mass, binary black hole scattering simulations (from \cite{Damour:2014afa}, \cite{Rettegno:2023ghr} and this work). For reference, we also present the corresponding values for the Lorentz factor $\gamma$, the Arnowitt-Deser-Misner (ADM) momenta $P_\mathrm{ADM}$ and the centre-of-mass velocity $v_\mathrm{cm}$.}
    \label{tab:NR_quantities}
    \setlength{\tabcolsep}{0.4cm}
    \begin{tabular}{c c c c}
    \hline
    \hline   
    $\Gamma$ & $\gamma$ & $P_\mathrm{ADM}$ & $v_\mathrm{cm}$\\
    \hline
    1.02264 & 1.09159 & 0.1145639 & 0.20926 \\
    1.04032 & 1.16453 & 0.1500000 & 0.27570 \\
    1.05548 & 1.22808 & 0.1750000 & 0.31995 \\
    1.07277 & 1.30167 & 0.2000000 & 0.36203 \\
    1.11346 & 1.47959 & 0.2500000 & 0.43979 \\
    1.16174 & 1.69928 & 0.3000000 & 0.50898 \\
    1.21688 & 1.96159 & 0.3500000 & 0.56981 \\
    \hline    
    \hline
    \end{tabular}
\end{table}

\subsection{Computational Setup}
The NR simulations were performed using the \texttt{Einstein Toolkit} \cite{EinsteinToolkit:2024_05}, an open-source NR code centered around the \texttt{Cactus} framework \cite{CactusURL}, with the assistance of \texttt{Simulation Factory} \cite{simfactory}. 
Post-processing of our simulation data was performed with \texttt{SimulationTools}, an open source Mathematica package \cite{SimulationTools}. 

Using the \texttt{McLachlan} thorn, we employ the BSSNOK formulation \cite{PhysRevD.52.5428, Baumgarte:1998te, Nakamura:1987zz} ($W$-variant \cite{Marronetti:2007wz}) of the 3+1 equations to perform time evolution of the system, supplemented with the following gauge choices: the moving punctures gauge condition \cite{Campanelli:2005dd, Baker:2005vv}, lapse evolution via $1+\log$ slicing \cite{Bona:1994dr}, and shift evolution using the hyperbolic gamma driver condition \cite{Alcubierre:2002kk}. 
We set the initial shift to zero and specify the initial data for the lapse profile using a pre-collapsed form \cite{Tichy:2003zg}
\begin{align}
    \alpha_\mathrm{HKV} = \frac{1 - \left( \frac{m_1}{2r_1} + \frac{m_2}{2r_2}\right)}{1 + \frac{m_1}{2r_1} + \frac{m_2}{2r_2}},
\end{align}
which is then averaged such that the initial lapse is given by
\begin{align}
    \alpha = \frac{1 + \alpha_\mathrm{HKV}}{2},
\end{align}
to enforce the condition $\alpha \in [0,1]$. 
The lapse, $\alpha_\mathrm{HKV}$ is physically motivated by the presence of an approximately helical killing vector (HKV) field and helps minimise the initial gauge dynamics \cite{Tichy:2003zg}. 
We use 8th-order finite differencing stencils with Kriess-Oliger artificial dissipation \cite{Kreiss:1973}. 
Adaptive mesh refinement is provided by the \texttt{Carpet} thorn, with the near zone around each black hole being computed using high resolution Cartesian grids. 
The Cartesian grids extend out to a radius of $100M$, at which point we transition to grids that are adapted to the spherical topology of the wave extraction zone, as implemented by the \texttt{Llama} thorn \cite{Pollney:2009yz}. 
The apparent horizons of the black holes are determined using the \texttt{AHFinderDirect} thorn \cite{Thornburg:2003sf} and the spin angular momenta of the black holes are calculated using the dynamical horizon formalism by the \texttt{QuasiLocalMeasures} thorn \cite{Dreyer:2002mx}. 

As in \cite{Damour:2014afa, Rettegno:2023ghr}, we seek initial data of the Bowen-York type \cite{Bowen:1980yu, Brandt:1997tf}, determined by the \texttt{TwoPunctures} thorn \cite{Ansorg:2004ds}. For binary black hole scattering scenarios, we have linear momentum of the form
\begin{align}
    \mathbf{P} = (P_x, P_y, P_z) = \pm \lvert \mathbf{P} \rvert \left( -\sqrt{1 - \left( \frac{b_\mathrm{NR}}{2X}\right)^2}, \frac{b_\mathrm{NR}}{2X}, 0\right),
\end{align}
where $b_\mathrm{NR}$ is the impact parameter (defined at a separation of $2X$) and we see that $P_\mathrm{ADM} = \lvert \mathbf{P} \rvert$. For consistency with \cite{Damour:2014afa, Rettegno:2023ghr}, we choose $X = 50M$ as a compromise between approximating infinite separation and computational expense. 

Since we only consider non-spinning black hole configurations, the ADM total angular momentum, $J_\mathrm{ADM}$, is given by
\begin{align}
    J_\mathrm{ADM} = L.
\end{align}
Here, $L$ is related to the impact parameter by
\begin{align}
    L = P_\mathrm{ADM}b_\mathrm{NR}.
\end{align}
Strictly speaking, the total ADM energy, $\Gamma_\mathrm{ADM}$ (and $J_\mathrm{ADM}$) includes unphysical junk radiation. We are only interested in the physical quantities, denoted the initial energy and initial angular momentum, which are given by
\begin{align}
    \Gamma_\mathrm{in} &= \Gamma_\mathrm{ADM} - \Gamma_\mathrm{junk} \approx \Gamma_\mathrm{ADM}, \\
    J_\mathrm{in} &= J_\mathrm{ADM} - J_\mathrm{junk} \approx J_\mathrm{ADM}.
\end{align}
In App.~\ref{Appendix: Junk}, we argue that the effects of junk radiation are negligible in our simulations. Subsequently, we equate the true initial quantities with the corresponding ADM values, and henceforth drop the subscript.

\subsection{NR Scattering Angle Extraction} \label{subsec:nr_scatter_angle_extract}
Following \cite{Damour:2014afa,Rettegno:2023ghr}, the motion of the black holes are tracked using the spherical polar coordinates $\lbrace r, \varphi \rbrace$ of the punctures in a center-of-mass frame. 
For each black hole, we measure the trajectories of the incoming $\varphi_{{\rm in}, i} (r)$ and outgoing $\varphi_{{\rm out}, i} (r)$ paths separately. 
The asymptotic angles $\varphi^{\infty}_{{\rm in/out},i}$ are determined by extrapolating a polynomial function of the inverse radius $u = 1/r$. 
The scattering angle for the $i$-th trajectory is defined to be
\begin{align}
\label{eq:nr_scatter}
\theta_i &= \varphi^{\infty}_{{\rm out},i} - \varphi^{\infty}_{{\rm in},i} - \pi. 
\end{align}
The fitting windows are slightly modified in comparison to \cite{Rettegno:2023ghr}, with the incoming domain corresponding to $r \in \left[ 20, 90 \right] M$ and the outgoing domain to $r \in \left[ 30, 180 \right] M$. 
We find a slightly larger domain helps stabilise the extrapolation, but note that the results are typically within the error bars quoted in \cite{Rettegno:2023ghr}. 
We adopt a least-squares algorithm that uses singular value decomposition (SVD) to drop singular values smaller than $10^{-12}$ times the maximum singular value \cite{Damour:2014afa,Rettegno:2023ghr}. 
We iteratively increase the polynomial order until there is no variation in the constant term up to the specified tolerance. 
The extrapolation errors are quoted using the maximum and minimum scattering
angle inferred across all polynomial orders, leading to dissymmetric bounds $\theta^{+ \delta_{+}}_{- \delta_{-}}$ \cite{Rettegno:2023ghr}. 
The relative scattering angle is defined as 
\begin{align}
\theta_{\rm NR} &= \frac{\theta_{1} + \theta_{2}}{2}.
\end{align}
The complete sequence of scattering angles extracted from our NR simulations are presented in ~\Crefrange{tab:nr_at_padm_0p114564}{tab:nr_at_padm_0p35} of App.~\ref{Appendix: NR Data}. 

\section{Relativistic Scattering Angles and the EOB Framework}
\label{Section: Relativistic Scattering Angles and the EOB Framework}

\subsection{Black Hole Recoil and the Relative Scattering Angle}

As a preface to our discussion on perturbative black hole scattering, we first provide a definition for scattering, including the role of radiative effects for generic unbound systems. Following this, we shall impose restrictions to reduce this general picture to that of our NR simulations. 

\begin{figure}
    \centering    
    \includegraphics[width=1\linewidth]{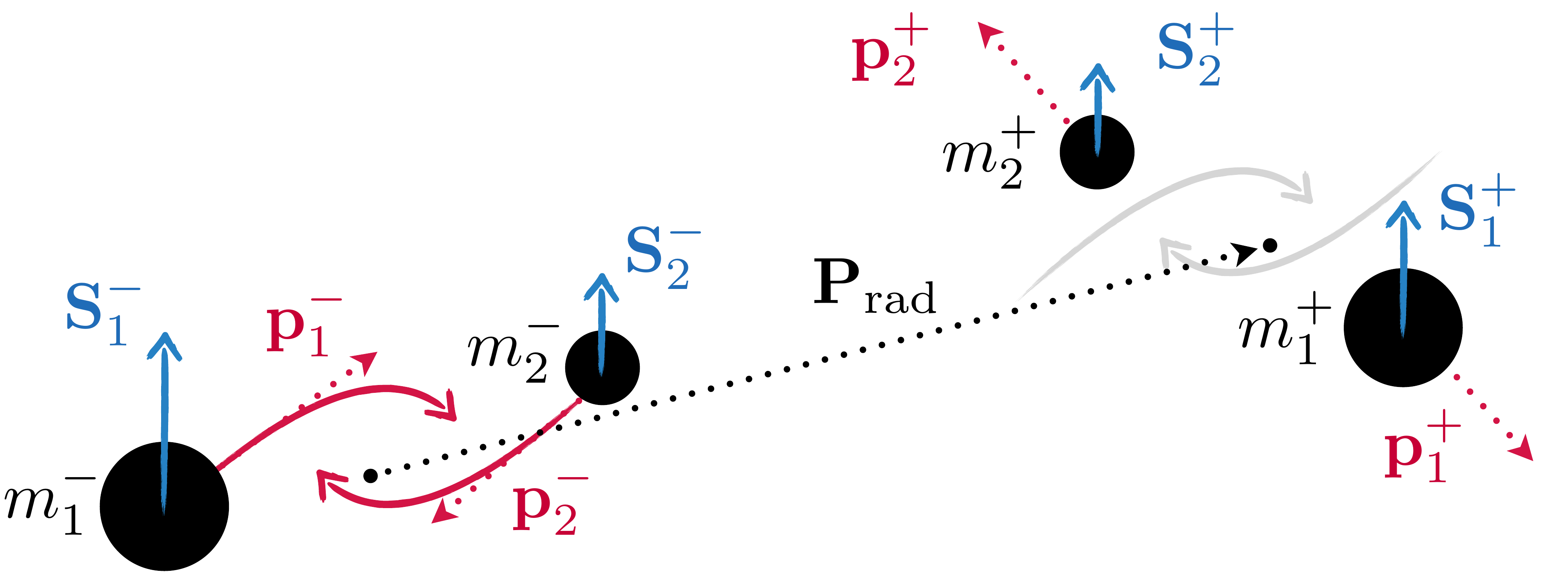}
    \caption{Effect of asymmetric radiation emission on an unbound black hole binary system, with $\mathbf{S}_i$ denoting the black hole spins. In our NR simulations, $\mathbf{S}_1 = \mathbf{S}_2 = 0$ and $m_1 = m_2$, so $\mathbf{P}_\mathrm{rad} = 0$.}
    \label{fig: bbh scatter}
\end{figure}

Intuitively, the scattering angle of each black hole can be defined as the angle between the ingoing and outgoing momenta of each black hole at spatial infinity, see Eq.~\ref{eq:nr_scatter} and Fig.~\ref{fig: bbh scatter}. 
In terms of the momenta $\mathbf{p}_i$, however, the scattering angle can equivalently be expressed as 
\begin{align}
\label{cons scatter angles}
    \cos(\theta_{i}) = \frac{\mathbf{p}_i^- \cdot \mathbf{p}_i^+ }{\lvert \mathbf{p}_i^- \rvert \lvert \mathbf{p}_i^+ \rvert} \;\;\; , \;\;\; i \in \{1,2\},
\end{align}
where the $-$ and $+$ states are to be interpreted as states at $\lbrace t \rightarrow -\infty, r \rightarrow \infty \rbrace$ and $\lbrace t \rightarrow \infty, r \rightarrow \infty \rbrace$ respectively. 
If one ignores radiative effects, the total relativistic energy and angular momentum are conserved. 
In this case, the impulse of each black hole (defined as $\Delta p_i^\mu = p_i^{\mu +} - p_i^{\mu -}$) obey $\Delta \mathbf{p}_1 = -\Delta \mathbf{p}_2$. 
Since $\mathbf{p}_1^{+/-} = -\mathbf{p}_2^{+/-}$, the scattering angle for each black hole is equal, i.e. $\theta_1 = \theta_2$. 

However, it is not sufficient to consider the conservative terms alone and we must incorporate dissipative effects.
Radiation-reaction angular momentum loss first enters at $\mathcal{O}(G^2)$ \cite{Damour:1981bh} with linear momentum loss entering at $\mathcal{O}(G^3)$ \cite{Kovacs:1977uw,Kovacs:1978uuw}. 
A key consequence is that the momentum impulses can generically differ such that 
\begin{align}
\Delta \mathbf{p}_1 + \Delta \mathbf{p}_2 \neq 0,
\end{align}
leading to a non-zero impulse of the system, known as recoil. If we can neglect the square of the recoil in the incoming center-of-mass frame, then we have that \cite{Bini:2021gat,Damour:2022ybd,Dlapa:2022lmu}
\begin{align}
\mathbf{p}_1^{+} + \mathbf{p}_2^{+} &= \mathbf{P}^{+} = - \mathbf{P}_{\rm rad}, \\
\mathbf{p}_1^{-} + \mathbf{p}_2^{-} &= \mathbf{P}^{-} = 0,
\end{align}
and we can introduce a relative scattering angle
\begin{align}
    \cos(\theta_\mathrm{rel}) = \frac{\mathbf{p}^{+} \cdot \mathbf{p}^{-}}{\lvert \mathbf{p}^{+} \rvert \lvert \mathbf{p}^{-} \rvert}.
\end{align}
Here, we have defined $\mathbf{p}^{-} = \mathbf{p}_1 = -\mathbf{p}_2$ and 
\begin{align}
    \label{momentum boost}
    \mathbf{p}^+ = \frac{E_2^+}{E_1^+ + E_2^+} \mathbf{p}_1^+ - \frac{E_1^+} {E_1^+ + E_2^+} \mathbf{p}_2^+ + \mathcal{O}(\mathbf{P}_{\mathrm{rad}}^2),
\end{align}
as outlined in \cite{Bini:2021gat,Damour:2022ybd,Dlapa:2022lmu,Buonanno:2024vkx}.

The radiative contributions to the scattering angle will take the generic form \cite{Bini:2012ji}
\begin{align}
\theta^{\rm rad} (E,J) &= - \frac{1}{2} \frac{ \partial \theta^{\rm cons}}{\partial E} E^{\rm rad} - \frac{1}{2} \frac{\partial \theta^{\rm cons}}{\partial J} J^{\rm rad},
\end{align}
where $\theta^{\rm cons}$ denotes the conservative contribution to the scattering angle and $\theta^{\rm rad}$ the radiative correction. 
The $\mathcal{O}(G^3)$ contribution has been calculated in \cite{Bern:2019nnu,Bern:2019crd,Kalin:2020fhe,Damour:2020tta,DiVecchia:2021ndb} and the $\mathcal{O}(G^4)$ term in \cite{Bini:2021gat,Bini:2022enm,Manohar:2022dea} for the odd-under-time-reversal contribution and in \cite{Dlapa:2022lmu,Bini:2022enm} for the even-under-time-reversal term. 

However, the NR simulations we present in section \ref{Section: Numerical Relativity} are non-spinning and equal-mass, so $\mathbf{P}_\mathrm{rad}$ vanishes and the relative scattering angle will coincide with the individual scattering angles defined in Eq. \ref{cons scatter angles}.
Maintaining consistency with the literature, the quantity of interest will be the relative scattering angle, which from this point forward we simply denote as $\theta$. 

\subsection{Scattering Angles in Post-Minkowskian Gravity}
The post-Minkowsian (PM) framework is a perturbative expansion of the Einstein field equations in terms of the gravitational coupling constant $G$, which only assumes that gravitational fields are weak, i.e. $[\epsilon \sim G M / (r c^2) \ll 1]$, and does not assume that the velocities are small with respect to the speed of light $c$ \cite{Blanchet:2013haa}. 
Schematically, the PM expansion can be written as a perturbative expansion around a Minkowski background
\begin{equation}
\label{metric perturbations}
    g_{\mu\nu} = \eta_{\mu\nu} + G h_{\mu\nu}^{(1)} + G^2 h_{\mu\nu}^{(2)} + \mathcal{O}(G^3),
\end{equation}
where $h_{\mu\nu}^{(i)}$ denotes the metric perturbation at $\mathcal{O}(G^i)$. 
As outlined in the introduction, a range of tools and techniques have been employed to generate the classical two-body scattering angle in PM gravity, which can be written as a PM-expansion in terms of the re-scaled orbital angular momentum $\ell$ 
\begin{equation} \label{PM scatter angle}
    \theta(\gamma, \ell) = \sum_{i} 2\frac{\theta_i(\gamma)}{\ell^i}.
\end{equation}
Alternatively, one can perform the expansion in terms of the impact parameter, $b = \Gamma M \ell / p_\infty$.
To obtain the scattering angle to $n$PM order, we simply take the partial sum of Eq.~\ref{PM scatter angle},
\begin{equation} 
    \label{eq:pm_expanded_angle}
    \theta_{n\mathrm{PM}}(\gamma, \ell) = \sum_{i=1}^n 2\frac{\theta_i(\gamma)}{\ell^i}.
\end{equation}

The coefficients $\theta_1(\gamma)$ and $\theta_2 (\gamma)$ can be calculated directly from the metric perturbations of Eq. \ref{metric perturbations} using position-space coordinates \cite{Portilla:1979xx,Westpfahl:1979gu}, see also \cite{Damour:2016gwp}. 
At the leading 1PM order, we have
\begin{align}
    \theta_1(\gamma) = \frac{2\gamma^2 - 1}{\sqrt{\gamma^2 - 1}},
\end{align}
and for the next-to-leading order term \cite{Damour:2017zjx}
\begin{align}
    \theta_2(\gamma) = \frac{3\pi}{8} \frac{(5\gamma^2 - 1)}{\Gamma(\gamma)}.
\end{align}
At order $\mathcal{O}(G^3)$ and beyond, the calculations become technically more involved and techniques commonly used in quantum field theory and scattering amplitudes have proven to be highly efficient at calculating results to high PM order.
The conservative 3PM scattering angle coefficient, $\theta_{3,\mathrm{cons}}(\gamma)$, was calculated in \cite{Bern:2019nnu, Kalin:2020fhe} together with a radiation reaction contribution at 3PM \cite{Damour:2020tta, DiVecchia:2021ndb}, $\theta_{3,\mathrm{rr}} (\gamma)$. The full 3PM contribution is given by
\begin{align}
    \theta_3 (\gamma) = \theta_{3,\mathrm{cons}} (\gamma) + \theta_{3,\mathrm{rr}} (\gamma).
\end{align}
At 4PM, the conservative scattering angle coefficient $\theta_{4,\mathrm{cons}}(\gamma)$ was computed in \cite{Dlapa:2021vgp, Bern:2021yeh} and the radiation reaction contribution, $\theta_{4,\mathrm{rr}}(\gamma)$, in \cite{Dlapa:2022lmu, Manohar:2022dea,Bini:2021gat,Bini:2012ji,Bini:2022enm}. The complete 4PM contribution to the scattering angle can be expressed as 
\begin{align}
    \theta_4 (\gamma) = \theta_{4, \mathrm{cons}} (\gamma) + \theta_{4, \mathrm{rr}} (\gamma)
\end{align}
where the radiation reaction can be further decomposed into time-odd and time-even contributions \cite{Dlapa:2022lmu}, i.e. $\theta_{4, \mathrm{rr}} (\gamma) = \theta_{4, \mathrm{rr}}^{\mathrm{odd}} (\gamma) + \theta_{4,\mathrm{rr}}^{\mathrm{even}} (\gamma)$. 

Recently, the conservative part of the 5PM scattering angle coefficient was computed in \cite{Driesse:2024xad} to first order in gravitational self-force (GSF). At the time of writing, the state-of-the-art PM-expanded scattering angle is given by
\begin{align}
    \theta_{5\mathrm{PM}, \mathrm{part}} (\gamma, \ell) = \sum_{i=1}^4 2\frac{\theta_i(\gamma)}{\ell^i} + 2\frac{\theta_{5, \mathrm{cons}}^\mathrm{1GSF} (\gamma)}{\ell^5},
\end{align}
where the subscript `part' indicates that we only have partial knowledge of the full 5PM term.

\subsection{Scattering Angles in the EOB Framework}
\label{SubSec: Scattering Angles EOB}
The EOB framework \cite{Buonanno:1998gg, Buonanno:2000ef, Damour:2000we} provides a mapping from the real two-body dynamics of a binary system to the dynamics of a single test body of mass $\mu$ moving in an effective spacetime metric. In the probe limit, for a test mass in a Schwarzschild or Kerr background, the symmetric mass ratio naturally serves as a deformation parameter. The real Hamiltonian is related to the effective Hamiltonian by the map \cite{Buonanno:1998gg, Buonanno:2000ef, Damour:2000we}
\begin{equation}
    H = M\sqrt{1 + 2\nu\left( \frac{H_\mathrm{eff}}{\mu} - 1 \right)}. 
\end{equation}
The effective energy of the system is equivalent to the effective Hamiltonian, i.e. $E_{\rm eff} = H_{\rm eff}$, and is related to the Lorentz factor by $H_{\rm eff} = \mu \gamma$, where $\gamma = - u^{-}_1 \cdot u^{-}_2$.

The EOB dynamics are described by a generalized mass-shell condition~\cite{Buonanno:1998gg,Damour:2000we}
\begin{equation}
    \mu^2 + g^{\mu\nu}_\mathrm{eff} P_\mu P_\nu + Q(X^\mu,P_\mu) = 0,
\end{equation}
where the first two terms describe the geodesic dynamics and $Q(X^\mu,P_\mu)$ encodes higher-order momentum terms beyond the quadratic contributions. 
Introducing re-scaled variables ($x^\sigma = X^\sigma/M$, $p_\sigma = P_\sigma/\mu, \hat{Q} = Q/\mu^2 $), the mass-shell condition can be written as
\begin{align}
    1 + g_\mathrm{eff}^{\mu\nu}p_\mu p_\nu + \hat{Q}(x^\mu,p_\mu) = 0.
\end{align}
In the non-spinning case, the effective metric has the general form 
\begin{align}
    g_{\mu\nu}^\mathrm{eff}dx^\mu dx^\nu = -A(r)dt^2 + B(r)dr^2 + C(r) d\Omega^2.
\end{align}
The symmetries of the mass-shell constraint lead to two constants of motion, $\ell = p_{\varphi}$ and $\gamma = - p_0$. The functional form of the radial momentum can be derived from the mass-shell constraint, leading to 
\begin{align} \label{Eq: radial action}
    p_r(\gamma, \ell, r) = \sqrt{\frac{B(r)}{A(r)}} \sqrt{\gamma^2 - A(r)\left(1 + \frac{\ell^2}{r^2} + \hat{Q}(r, \gamma) \right)}.
\end{align}
Alternatively, this expression can be inverted to obtain the effective Hamiltonian \cite{Buonanno:1998gg},
\begin{align}
    H_{\rm eff} &= \sqrt{A(r) \left( 1 + \frac{\ell^2}{r^2} + \frac{p^2_r}{B(r)} + \hat{Q}(r,\gamma) \right)}.
\end{align}
From Hamilton-Jacobi theory, the concomitant scattering angle is given by \cite{Damour:2016gwp}
\begin{equation} \label{Eq: Hamilton-Jacobi Scattering Angle}
    \theta(\gamma, \ell) = -\pi - \displaystyle\int_{-\infty}^{\infty} \mathrm{d}r \frac{\partial p_r (\gamma, \ell, r)}{\partial \ell},
\end{equation}
where the limits of integration should be interpreted as follows \cite{Damour:2017zjx}: $-\infty$ refers to the incoming state $\lbrace t \rightarrow -\infty, r \rightarrow \infty \rbrace$, whereas $\infty$ refers to the outgoing state $\lbrace t \rightarrow \infty, r \rightarrow \infty \rbrace$. 

Whilst the scattering angle $\theta$ is gauge invariant, the effective potential will explicitly depend on the functional form of the metric potentials $\lbrace A, B \rbrace$ and the non-geodesic term $Q$. This includes gauge choices, resummation, and any additional information that can be incorporated into the model, such as analytical post-Newtonian (PN) or PM calculations or information from NR simulations. 

\section{Comparisons to NR}
\label{Section: Comparisons to NR}
Much like the PN expansion \cite{Blanchet:2013haa}, it has been shown that the non-resummed PM-expanded scattering angles demonstrate poor convergence towards NR \cite{Damour:2022ybd,Rettegno:2023ghr}. 
This motivates the exploration of \textit{resummation} strategies and attempts to bound the order to which we need PM information across the parameter space.  
In the following, we systematically explore a range of analytical approaches to defining the scattering angle, with comparisons against our suite of NR simulations. 
In particular, we focus on the approach to the high-energy limit for equal-mass non-spinning binaries, which proves to be challenging for all resummation schemes considered. 
The methods we focus on include: i) the non-resummed PM-expansion of the scattering angle, ii) the $\mathcal{L}$-resummation proposed in \cite{Damour:2022ybd}, iii) the $w^{\rm eob}$ resummation of the PM-expanded scattering angles using the EOB framework proposed in \cite{Damour:2022ybd}, and iv) the $w^{\rm SEOB-PM}$ resummation introduced in \cite{Buonanno:2024byg} based on the SEOBNRv5 Hamiltonian \cite{Khalil:2023kep}. Note that we do not aim to exhaustively survey all attempts at resummation of PM information, such as the procedure explored in \cite{Kalin:2019rwq,Kalin:2019inp}, corresponding to a resummation of the one-loop contributions at $n$PM.

\subsection{The Post-Minkowskian (PM) Expansion}
\label{sec:PMexp}
In the first approach, we explore the behaviour of the non-resummed PM-expanded scattering angle, as given by Eq.~\ref{eq:pm_expanded_angle}. 
A priori, we know that the scattering angles obtained from a non-resummed PM expansion show slow convergence to NR with increasing PM order, e.g. \cite{Damour:2022ybd, Rettegno:2023ghr}. 
While the PM expansion agrees with NR predictions in the weak-field (high angular momentum) limit, significant discrepancies start to arise as the system approaches the strong-field (low angular momentum) regime.

\begin{figure*}
  \centering
  \subfloat{
      \includegraphics[width=.5\textwidth]{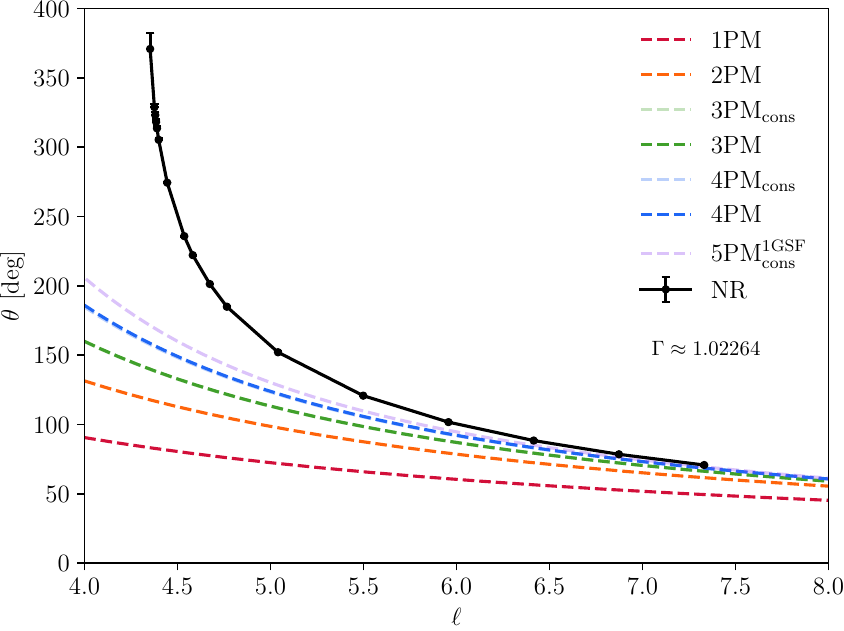}
  }
  \subfloat{
      \includegraphics[width=.5\textwidth]{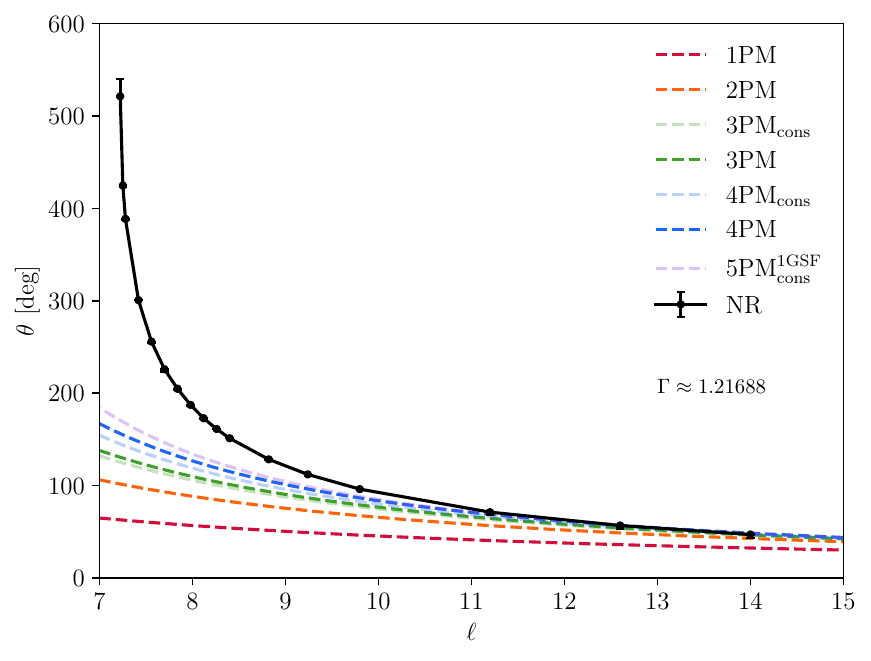}
  }    
  \caption{Predictions of the scattering angle from NR data against the PM-expanded scattering angles of Eq. \ref{eq:pm_expanded_angle}. Left: Comparison at the lowest available energy, $\Gamma_1$. Right: Comparison at the highest available energy, $\Gamma_7$. At all available energies, the agreement in the strong-field regime is very poor.
  }
  \label{fig:pm_expanded_vs_nr_plots}
\end{figure*}

In Fig.~\ref{fig:pm_expanded_vs_nr_plots}, we compare the PM-expanded scattering angle with the lowest and highest energy NR simulations in our dataset.
As could be expected (see ~\cite{Damour:2022ybd,Rettegno:2023ghr}), the PM-expanded scattering angles are insufficient at all energies available in our suite of simulations.
Increasing the PM order yields noticeable improvement in the agreement with NR, but still fails to accurately capture the plunge. Since only conservative information is available up to 1GSF at 5PM, we also present the conservative predictions for each PM order, noting that there are no radiative contributions at 1PM and 2PM. 
For the conservative predictions, we consistently discard \textit{all} radiative contributions at all PM orders. 
Including partial 5PM information results in a noticeable improvement over the full 4PM result. This highlights the utility of higher-order PM terms in achieving more accurate strong-field predictions when using non-resummed PM-expanded scattering angles and aptly demonstrates the slow convergence of the PM series.

Naively, we would expect the scattering angles and related quantities to admit a finite limit at a given PM order.  
However, recent work has identified the existence of power-law divergences at 4PM \cite{Damour:2020tta,Bini:2022enm,Dlapa:2022lmu}, for example $\theta_4 / \gamma^4 \propto \gamma^{1/2}$ in the high-energy limit \cite{Bini:2022enm,Damour:2022ybd}.
See also the related discussion in \cite{Gruzinov:2014moa,Ciafaloni:2015xsr,Herrmann:2021tct,DiVecchia:2022nna} and the presence of logarithmic mass singularities identified in \cite{Dlapa:2022lmu}.
Furthermore, we see divergent properties of the 5PM conservative (to 1 GSF order) term in both the low and high energy limits.
The 3PM term, on the other hand, demonstrates comparatively good behaviour in the high-energy limit \cite{Amati:1990xe,Damour:2020tta,DiVecchia:2021ndb,Damour:2022ybd}, though we find the agreement with NR is still relatively poor. 
A related issue is the apparent noncommutativity of the PM results in the limits $G \rightarrow 0$ and $\gamma \rightarrow \infty$ \cite{Damour:2020tta}. 
In particular, \cite{Damour:2020tta} outlined a proof that the dynamics undergoes a transmutation of PM order in which terms of order $\mathcal{O}(G^{\geq 4})$ are reduced to $\mathcal{O}(G^3 \ln G)$ as $\gamma \rightarrow \infty$.
These issues raise a number of potential challenges in working with analytical PM results. 

Finally, as computations to 6PM order and higher are expected to be technically very challenging, it may be difficult for non-resummed PM information to provide reliable strong-field predictions in the very near future.
It is imperative to stress that the incorporation of high-order PM information in a resummed form will still be a critical component.  
A second key issue is the inability of PM-expanded results to accurately predict the critical angular momentum that marks the transition from scatter to plunge. 
Consistent with previous studies, the poor convergence of the PM scattering angles emphasises the need to develop \textit{resummation} strategies for incorporating high-order PM information, particularly in the context of source modeling \cite{Damour:2022ybd,Rettegno:2023ghr,Buonanno:2024byg,Buonanno:2024vkx}.

\subsection{$\mathcal{L}$-Resummation}\label{Section: L-Resummed Comparison}

\subsubsection{The $\mathcal{L}$-Resummation of the Scattering Angle}
A resummation strategy was recently proposed in \cite{Damour:2022ybd}, leveraging the presence of a logarithmic divergence in the leading-order geodesic limit. 
This behaviour arises due to the coalescence of the two largest positive real roots of Eq. \ref{Eq: radial action} in the limit $\ell \rightarrow \ell_0^+$. 
This is assumed to be universal and general to all PM orders, with a singular structure of the form~\cite{Damour:2022ybd}
\begin{align}
\theta (\ell) \overset{\ell \rightarrow \ell^{+}_0}{\sim} \frac{\ell}{\ell_0} \ln \left[ \frac{\ell}{\ell - \ell_0} \right].
\end{align}
Noting that this can be re-expressed as a formally convergent power-series, \cite{Damour:2022ybd} introduced a function $\mathcal{L}(x)$ to capture the singular behaviour,
\begin{align}
\mathcal{L}(x) &= \frac{1}{x} \ln \left[ \frac{1}{1-x} \right],
\end{align}
valid when $|x| < 1$. 
This allows one to define a $\mathcal{L}$-resummation of the PM-expanded scattering angle by factoring out the singular term \cite{Damour:2022ybd}, i.e.
\begin{align} 
\label{Eq: L-resummed}
    \theta_{n\mathrm{PM}}^\mathcal{L}(\gamma, \ell; \ell_0) = \mathcal{L}\left( \frac{\ell_0}{\ell} \right) \hat{\theta}_{n\mathrm{PM}}(\gamma, \ell; \ell_0),
\end{align}
where
\begin{align}
    \hat{\theta}_{n\mathrm{PM}}(\gamma, \ell; \ell_0) = \sum_{i=1}^{n} 2\frac{\hat{\theta}_i(\gamma; \ell_0)}{\ell^i}.
\end{align}
The $n$th term in the large-$\ell$ expansion of $\theta_{n\mathrm{PM}}^\mathcal{L}$ is matched to that of $\theta_{\rm nPM}$, uniquely determining the coefficients $\hat{\theta}_i$. 
The relations between these coefficients are specified up to 6PM in App.~\ref{app:pm_to_resummed_map}.

As the $\mathcal{L}$-resummed model explicitly depends on the critical angular momentum $\ell_0$, one needs to \textit{a-priori} calculate this value. 
In \cite{Damour:2022ybd}, Cauchy's rule was used to analytically estimate $\ell_0$ from perturbative PM information, in which the PM expansion of $\mathcal{L}(\ell_0 / \ell)$ can be compared order-by-order to that of $[ \theta_{n\mathrm{PM}}(\gamma, \ell)/\theta_{1\mathrm{PM}}(\gamma, \ell) ]$. By applying Cauchy's rule, the analytical estimate for $\ell_0$ is then given by \cite{Damour:2022ybd}
\begin{align}\label{Eq. cauchy_estimate}
    \ell_0^{n\mathrm{PM}}(\gamma) = \left[ n \frac{\theta_n(\gamma)}{\theta_1(\gamma)} \right]^{\frac{1}{n-1}} ,
\end{align}
where $n$ denotes the PM order.
The most accurate estimate of $\ell_0$ should be provided by the highest-known PM term. 
Inserting this value of $\ell_0$ into Eq.~\ref{Eq: L-resummed} completes the $\mathcal{L}$-resummed model. 
Alternatively, one may use NR simulations, which contain non-perturbative physics, to estimate $\ell_0$ or calibrate higher-order PM terms that are as-of-yet unknown or otherwise incomplete.

This resummation scheme was recently extended to incorporate information about the self-force term of the singular behaviour \cite{Long:2024ltn}, demonstrating a stronger divergence than the term derived from the geodesic-limit alone. 
Whilst this scheme has only been applied to a scalar-field model, it is expected that the resummation technique can significantly enhance the validity of PM calculations of black hole scattering for particles subject to the full gravitational self-force.

\begin{figure*}
  \centering
  \subfloat{
      \includegraphics[width=.5\textwidth]{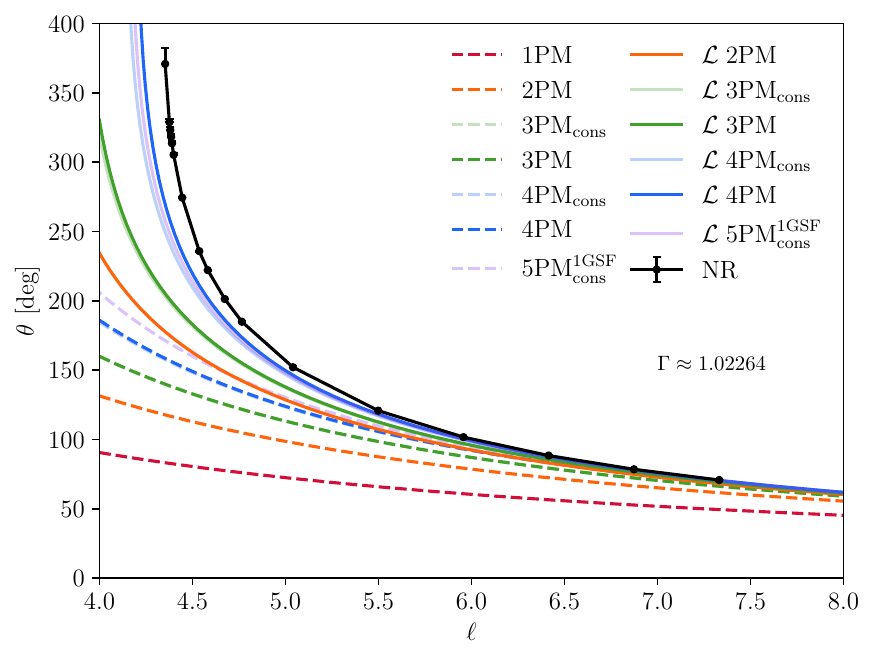}
  }
  \subfloat{
      \includegraphics[width=.5\textwidth]{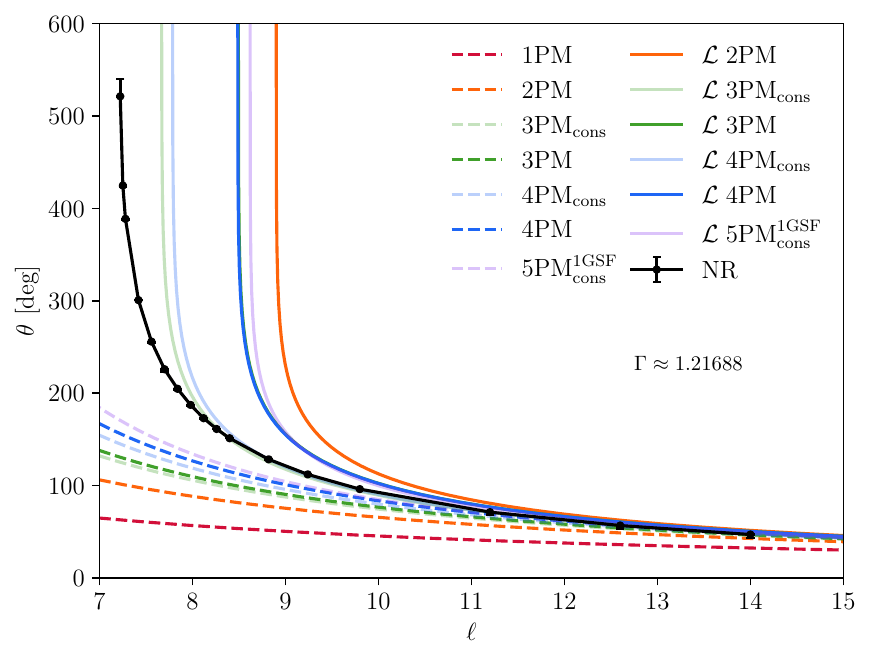}
  }    
  \caption{Comparison of the $\mathcal{L}$-resummed scattering angles with the scattering angles extracted from our NR simulations. Left: Comparison at the lowest available energy, $\Gamma_1$. Right: Comparison at the highest available energy, $\Gamma_7$. We supplement both comparisons with the predictions of the PM-expanded scattering angles for reference.
  }
  \label{fig:L_resummed_vs_nr_plots}
\end{figure*}

\begin{figure}
    \centering
    \includegraphics[width=1\linewidth]{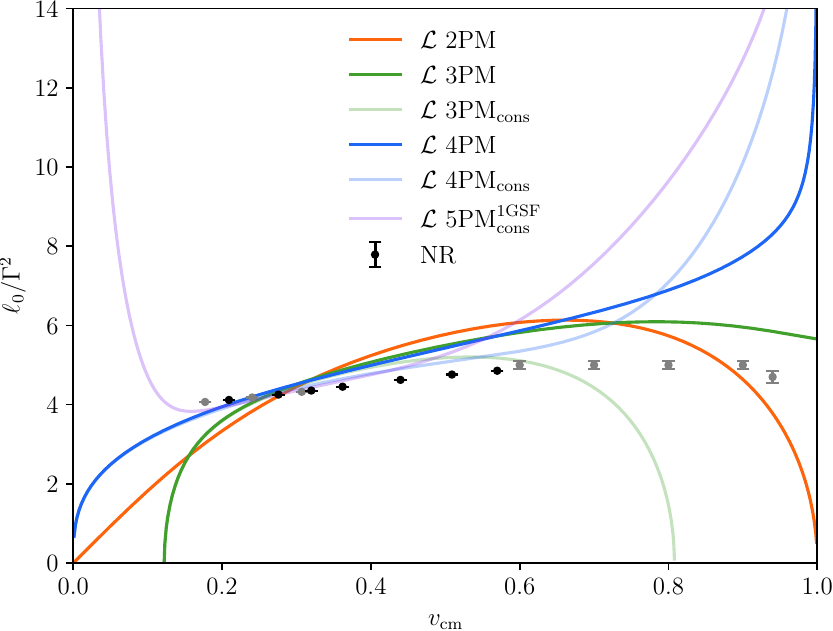}
    \caption{Re-scaled Cauchy estimate of $\ell_0$ (Eq. \ref{Eq. cauchy_estimate}) compared against the value extracted from our NR data as obtained following a similar procedure to \cite{Damour:2022ybd}. Grey data points are calculated/taken from \cite{Albanesi:2024xus, Shibata:2008rq, Sperhake:2009jz}.}
    \label{fig:L_resummed_critical_angular_momentum}
\end{figure}

\subsubsection{Performance of the $\mathcal{L}$-Resummation} \label{Sec. Performance of the L Resummation}
In \cite{Damour:2022ybd}, the $\mathcal{L}$-resummed model demonstrated significant improvements in accuracy compared to NR data at each successive PM order. 
However, the analysis was limited to a single fixed energy, $\Gamma_1$.
We expand on \cite{Damour:2022ybd} in a few directions. 
First, we extend the baseline model to include the recently derived 5PM conservative contribution (at 1GSF order), which, as we will show, offers a slight improvement over the 4PM conservative estimate at low energies. 
Second, we investigate the energy dependence of the $\mathcal{L}$-resummation scheme. 
Unless otherwise specified, we assume that $\ell_0$ is calculated from the Cauchy estimate in Eq.~\ref{Eq. cauchy_estimate}. 
The main results of our comparison are shown in Fig.~\ref{fig:L_resummed_vs_nr_plots}, with additional plots provided in Fig.~\ref{Fig: Additional Lresum_plots} of App.~\ref{Appendix: Additional Lresum_plots}.

For low energies, the PM hierarchy is not robust. 
At the lowest energy, the order is broadly as expected with the complete 4PM term being the most-consistent with the NR data. 
For the second-lowest energy, the 2PM term is under-attractive whereas the 3PM term is over-attractive. 
The inclusion of the 4PM term surprisingly reduces the accuracy of the model.
As we approach higher energies, the hierarchy of PM orders is largely preserved with each successive PM order typically improving the accuracy with respect to NR. 
The radiative contributions become increasingly problematic, as per the discussion in Sec.~\ref{sec:PMexp}. 
The behaviour as a function of energy is governed by the non-trivial structure of the scattering angles at each iterative PM order as well as the estimate of the critical angular momentum. 

Based on the above discussion, it is important to explore the PM-based Cauchy estimate of the critical angular momentum across the full range of energies. 
In particular, we can use the NR simulations to assess the accuracy of the analytical predictions and gauge the relative importance of higher-order information in the estimate of $\ell_0$.
First, we define the centre-of-mass velocity of the system to be
\begin{align}
    v_\mathrm{cm} = \sqrt{\frac{\gamma - 1}{\gamma + 1}},
\end{align}
such that $v_\mathrm{cm} \in (0,1), \, \forall \, \gamma \in (1, \infty)$. 
In Fig.~\ref{fig:L_resummed_critical_angular_momentum}, we show how the analytical prediction of $\ell_0$ behaves as a function of $v_\mathrm{cm}$ for all available PM orders. 
We include, alongside the NR data shown in Fig.~\ref{fig:all_nr_data}, the results of additional NR simulations in grey (from left to right: the first 3 data points were calculated using data from \cite{Albanesi:2024xus}, the following 4 points were computed in \cite{Shibata:2008rq}, and the last data point was obtained in \cite{Sperhake:2009jz}). 

Whilst the hierarchical structure is approximately in line with our naive expectations for the lowest energy, $\Gamma_1$, we can clearly see how the hierarchy is evolving with energy and how the PM expansion fails to consistently converge towards to the NR values. 
This is particularly evident in the low- and high-velocity limits.
Despite naive expectations, the prior discussion on the divergences present in the PM coefficients is in agreement with the observed behaviour, e.g. \cite{Damour:2022ybd}.
The power-law divergence in the 4PM coefficient induces divergences in the high-energy behavior of the 4PM and higher-order terms. 
In contrast, the well-behaved high-energy limit of the 3PM term helps provide a reasonably robust estimate of the critical angular momentum but is still inaccurate with respect to NR. 
In the limit $v_\mathrm{cm} \rightarrow 0$, the estimate at the partial 5PM level shows a strong divergence,
in contrast to the lower-order PM estimates. 
In particular, the Cauchy estimate for 4PM behaves as $\propto (\gamma - 1)^{1/6}$ \cite{Damour:2022ybd}, while the 3PM estimate exhibits a branch-cut singularity associated to the change in sign of $\theta_3$ \cite{Damour:2022ybd}. 
It would be interesting to explore this regime with dedicated NR simulations to further understand the low-energy behaviour of these functions.

So long as the PM terms exhibit divergences, the $\mathcal{L}$-resummation scheme will remain unreliable at high energies \cite{Damour:2022ybd}. 
The anticipated non-trivial structure of higher-order PM terms makes it difficult to predict how they will impact the $\theta_{n\mathrm{PM}}^\mathcal{L}$ expansion and hence how they will mitigate against the above behavior.
For example, if the 5PM informed models are to accurately reproduce the numerical results across all energies, the 2GSF and radiative corrections would have to counter the divergences present in the 4PM and partial 5PM terms. 

We now focus on different approaches to incorporating numerical information into the $\mathcal{L}$-resummation framework. 
We construct a general template of the form \cite{Damour:2022ybd}
\begin{multline} \label{eq:L_resummed_template}
    \theta_{5\mathrm{PM},\mathrm{X}}^\mathcal{L}(\gamma, \ell; \ell_{0,\mathrm{X}}, \theta_{5, \mathrm{X}})  = \mathcal{L}\left(\frac{\ell_{0, \mathrm{X}}}{\ell}\right) \times \\ \Bigg[ \Bigg. \hat{\theta}_{4\mathrm{PM}}(\gamma, \ell; \ell_{0, \mathrm{X}}) + 2\frac{\hat{\theta}_{5}(\gamma; \ell_{0, \mathrm{X}}, \theta_{5, \mathrm{X}})}{\ell^5} \Bigg. \Bigg],
\end{multline}
where
\begin{align} \label{eq:L_resummed_template_coefficient}
    \hat{\theta}_{5}(\gamma; \ell_{0, \mathrm{X}}, \theta_{5, \mathrm{X}}) &= \theta_{5, \mathrm{X}} - \frac{\ell_{0, \mathrm{X}}}{2}\theta_4(\gamma) - \frac{\ell_{0, \mathrm{X}}^2}{12}\theta_3(\gamma) \\ \nonumber &\; - \frac{\ell_{0, \mathrm{X}}^3}{24}\theta_2(\gamma) - \frac{19\ell_{0, \mathrm{X}}^4}{720}\theta_1(\gamma).
\end{align}
A template of this form allows us to simultaneously fit higher-order PM coefficients as well as the critical angular momentum $\ell_0$. 
We introduce the placeholder $X \in \{\mathrm{I}, \mathrm{II} \}$ to denote the type of fit performed. 
In the first scheme (I), we only fit for $\theta_5$ with the critical angular momentum determined using the Cauchy estimate from Eq. \ref{Eq. cauchy_estimate}, i.e.
\begin{align} \label{eq:analytical_estimate_of_ell_template}
    \ell_{0,\mathrm{I}}(\gamma) = \left[ 5 \frac{\theta_{5,\mathrm{I}}}{\theta_1(\gamma)} \right]^{\frac{1}{4}}.
\end{align}

\begin{figure}[h]
    \centering
    \includegraphics[width=1\linewidth]{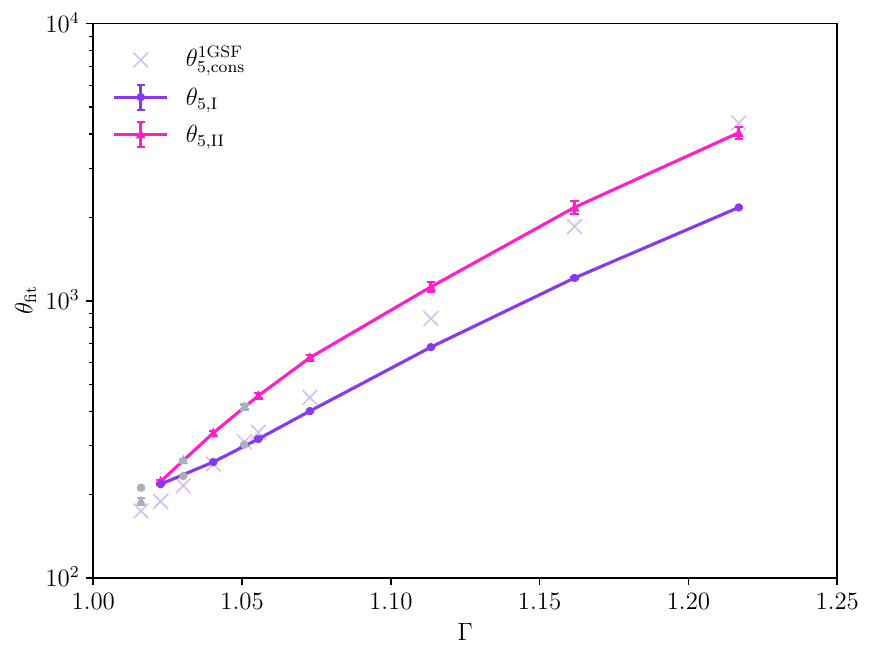}
    \caption{Values of $\theta_{5,\mathrm{I}}$ and $\theta_{5,\mathrm{II}}$ at energies $\Gamma_1$-$\Gamma_7$, compared with the values of $\theta_{5, \mathrm{cons}}^{1\mathrm{GSF}}$. Fits performed against data from \cite{Albanesi:2024xus} are overlaid in grey.}
    \label{fig:L_resummed_fit_value}
\end{figure}
In the second scheme (II), we treat both $\ell_0$ and $\theta_{5}$ as free coefficients to be inferred from the data. 
The first scheme allows us to explore the impact of incorporating higher-order PM information in the $\mathcal{L}$-resummation. The second scheme allows us to explore the sensitivity of the framework to the accuracy with which we estimate the critical angular momentum and the impact of an NR-calibrated estimate. 
We re-calculate the scattering angles using both fitting schemes, and compare the results with NR data. 

\begin{figure*}[htp]
  \centering
  \subfloat{
      \includegraphics[width=.5\textwidth]{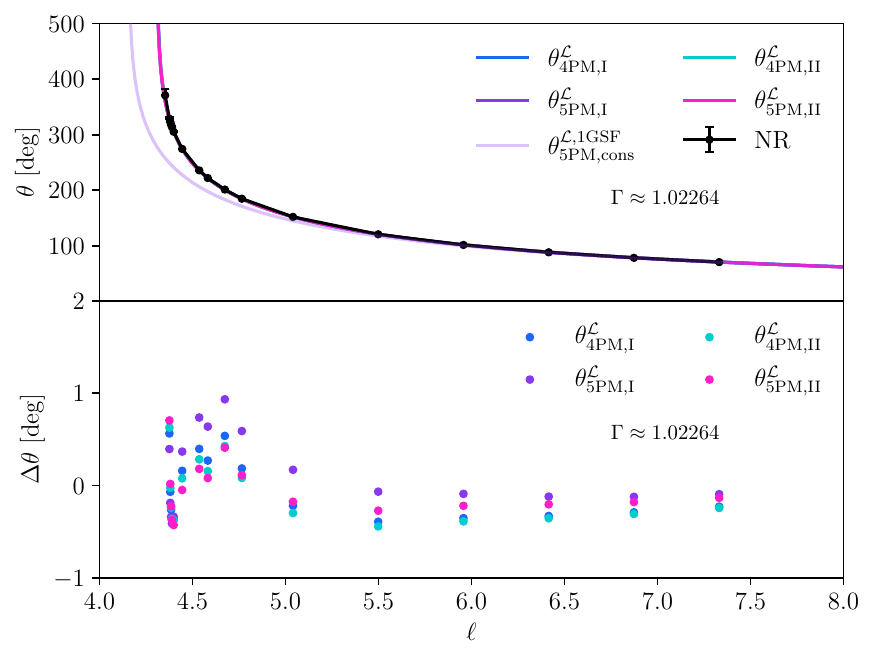}
  }
  \subfloat{
      \includegraphics[width=.5\textwidth]{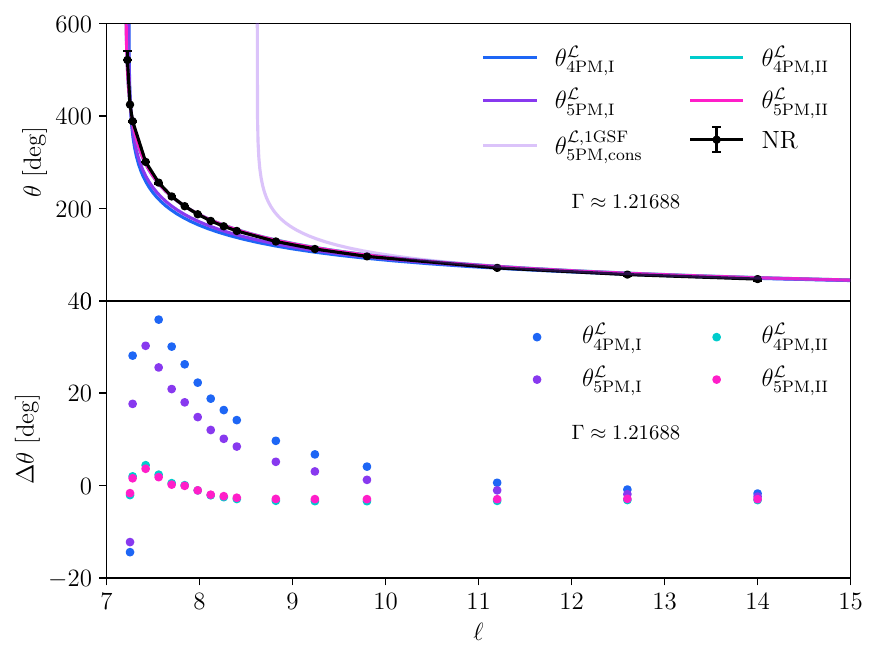}
  }    
  \caption{Analysis of the predictions of $\theta_{4/5\mathrm{PM},\mathrm{I}}^\mathcal{L}$ and $\theta_{4/5\mathrm{PM},\mathrm{II}}^\mathcal{L}$. For comparison, we plot these alongside $\theta_{5\mathrm{PM},\mathrm{cons}}^{\mathcal{L}, 1\mathrm{GSF}}$. Additionally, we show the residuals against the NR data and note that the residuals grow monotonically with energy.}
  \label{Fig: L-Resummed Fit Analysis}
\end{figure*}

In Fig.~\ref{fig:L_resummed_fit_value}, we show a direct comparison between $\theta_{5, \mathrm{cons}}^{1\mathrm{GSF}}(\gamma)$ and $\theta_{5, \mathrm{I/II}}$. We see that both sets of NR fits are of the same order of magnitude as $\theta_{5, \mathrm{cons}}^{1\mathrm{GSF}}(\gamma)$ across the range of energies. At $\Gamma_1$, the values of $\theta_{5, \mathrm{I}}$ and $\theta_{5, \mathrm{II}}$ are very similar, differing by $\sim 2-3\%$. 
As shown in Fig. \ref{Fig: L-Resummed Fit Analysis}, $\theta_{5\mathrm{PM},\mathrm{I}}^\mathcal{L}$ and $\theta_{5\mathrm{PM},\mathrm{II}}^\mathcal{L}$ give almost identical predictions for the scattering angle, with maximum residuals relative to NR being on the order of $\sim 1^\circ$. 
As the Cauchy estimate of the critical angular momentum is quite reasonable, being comparable to the NR-informed values, higher-order PM information is likely driving the improvement.

Beyond $\Gamma_1$, the predictions of $\theta_{5\mathrm{PM},\mathrm{II}}^\mathcal{L}$ agree with the NR data, while the performance of $\theta_{5\mathrm{PM},\mathrm{I}}^\mathcal{L}$ steadily degrades as we increase the energy. 
By the time we reach the highest energy, a number of intriguing features become apparent. 
We now have a visible difference between $\theta_{5\mathrm{PM},\mathrm{I}}^\mathcal{L}$ and $\theta_{5\mathrm{PM},\mathrm{II}}^\mathcal{L}$, with maximum residuals from NR $\sim 30\%$ and $\sim 4\%$, respectively. Interestingly, we find the (upper limit) value of $\theta_{5, \mathrm{II}}$ is within $\sim 4\%$ of $\theta_{5, \mathrm{cons}}^{1\mathrm{GSF}}(\gamma)$, enabling a direct comparison of $\ell_0^{5\mathrm{PM}}$ and $\ell_{0, \mathrm{II}}$. Comparison of $\theta_{5\mathrm{PM}, \mathrm{II}}^\mathcal{L}$ and $\theta_{5\mathrm{PM}, \mathrm{cons}}^{\mathcal{L}, 1\mathrm{GSF}}$ in Fig.~\ref{Fig: L-Resummed Fit Analysis} shows how an improved prediction of $\ell_0$ can enhance the predictions of the $\mathcal{L}$-resummation, whether that is calibrating $\ell_0$ to NR or via a new and improved analytical construction.

Additionally, we analyse $\theta_{4\mathrm{PM},\mathrm{I/II}}^\mathcal{L}$, a 4PM analogue of the fitting procedure in Eqs. \ref{eq:L_resummed_template} and \ref{eq:L_resummed_template_coefficient}. 
At $\Gamma_1$, the residuals of each fit are of the same order, with maximum values $\sim 1^\circ$. Across $\Gamma_2$-$\Gamma_7$, we see that $\theta_{5\mathrm{PM},\mathrm{I}}^\mathcal{L}$ outperforms $\theta_{4\mathrm{PM},\mathrm{I}}^\mathcal{L}$, giving further evidence in favour of higher order PM information. We also find the predictions of $\theta_{5\mathrm{PM},\mathrm{II}}^\mathcal{L}$ have comparable accuracy when compared to $\theta_{4\mathrm{PM},\mathrm{II}}^\mathcal{L}$, with both outperforming the $\theta_{4/5\mathrm{PM},\mathrm{I}}^\mathcal{L}$ fits. 
This indicates that an accurate estimation of the critical angular momentum $\ell_0$ may have greater importance than including higher-order PM information in the $\mathcal{L}$-resummed scheme. 

An important caveat to the above discussion regards the physical interpretation of our \textit{pseudo}-PM terms. 
When calibrating such coefficients against NR data, these terms can absorb partial contributions from effective higher-order terms as they attempt to approximate the full non-perturbative information contained in the NR simulations. 
The physical interpretation of the pseudo-PM terms should be treated with caution, as they may reflect effective contributions beyond their nominal perturbative order.

\subsection{The $w^{\rm eob}$ Model}
\subsubsection{The $w^\mathrm{eob}$ Resummation of the Scattering Angle} \label{SubSec. weob}

The $w^\mathrm{eob}$ model was first proposed in \cite{Damour:2022ybd}, as a reformulation of the scattering angle in terms of a PM-expanded EOB potential. In this model, one adopts the post-Schwarzschild (PS) gauge \cite{Damour:2017zjx} in which the metric potentials are fixed to those of Schwarzschild, with $\hat{Q}(r,\gamma)$ admitting a PM expansion. It is convenient to choose isotropic coordinates such that $g_\mathrm{Schw}dx^\mu dx^\nu \rightarrow \bar{A}(\br)dt^2 + \bar{B}(\br)d\bar{x}^id\bar{x}^i$. In these coordinates, the effective Hamiltonian reads
\begin{align}
    H_{\rm eff} &= \sqrt{\bar{A}(\br) \left( 1 + \frac{\ell^2}{ \bar{B}(\br) {\br}^2 } + \frac{ p^2_{\br} }{ \bar{B} (\br) } + \hat{Q}(\br,\gamma) \right)},
\end{align}
leading to the mass-shell condition 
\begin{align}
    p_{\br}(\gamma, \ell, \br) = \sqrt{p_\infty^2 + w^\mathrm{eob}(\br, \gamma) - \frac{\ell^2}{\br^2}},
\end{align}
where
\begin{align}
    w^\mathrm{eob}(\br, \gamma) &= \gamma^2 \left( \frac{\bar{B}(\br)}{\bar{A}(\br)} - 1 \right) + 1 \\ \nonumber &\qquad \qquad \qquad  - \bar{B}(\br) - \bar{B}(\br)\hat{Q}(\br, \gamma).
\end{align}

In the PS gauge, $\hat{Q}(\br, \gamma)$ is expressed as a PM expansion in the inverse of the radius $\br (= \bar{R}/GM)$
\begin{align}
    \hat{Q}(\br,\gamma) = \sum_{i = 2} \frac{\hat{Q}_i(\gamma)}{\br^i},
\end{align}
such that $\br^{-i} \propto G^i$ serves to count the PM order. 
It was shown in \cite{Damour:2016gwp} that up to $\mathcal{O}(G)$, the two-body dynamics are described by the effective Schwarzschild metric. 
Hence, the expansion starts at $\mathcal{O}(G^2)$. The relation between $w^\mathrm{eob}(\br, \gamma)$ and $\hat{Q}(\br, \gamma)$ allows $w^\mathrm{eob}(\br, \gamma)$ to also take the PM-expanded form \cite{Damour:2017zjx}
\begin{align}
    w^\mathrm{eob}(\br,\gamma) = \sum_{i} \frac{w_i(\gamma)}{\br^i},
\end{align}
where we now have terms at $\mathcal{O}(G)$ due to contributions from the Schwarzschild metric potentials. As before, we define $w^\mathrm{eob}$ to $n\mathrm{PM}$ order as the partial sum
\begin{align}
    w^\mathrm{eob}_{n\mathrm{PM}}(\br,\gamma) = \sum_{i=1}^n \frac{w_i(\gamma)}{\br^i}.
\end{align}
This PM expansion of the EOB potentials will be denoted the EOB-PM potential.
Using Eq. \ref{Eq: Hamilton-Jacobi Scattering Angle}, the scattering angle is given by
\begin{align} \label{Eq: weob Scattering Angle}
    \theta(\gamma, \ell) = -\pi + 2\ell \int_{\brmin (\gamma, \ell)}^{\infty} \frac{\mathrm{d} \br}{\bar{r}^2 \sqrt{p_\infty^2 + w^\mathrm{eob}(\gamma, \bar{r}) - \frac{\ell^2}{\bar{r}^2}}},
\end{align}
where $\brmin (\gamma, \ell)$ is defined as the largest real, positive root of $p_{\br}$.
The main idea behind the $w^\mathrm{eob}$ model is to determine $w^\mathrm{eob}_{n\mathrm{PM}}$ (up to the desired PM order) and allow this to replace the full EOB potential, $w^\mathrm{eob}$, in Eq. \ref{Eq: weob Scattering Angle}. In \cite{Damour:2019lcq}, it was shown that one can determine the $w_i(\gamma)$ coefficients by inserting $w^\mathrm{eob}_{n\mathrm{PM}}$ into Eq. \ref{Eq: weob Scattering Angle}, performing a PM expansion of the integrand (expansion in $\ell$) and take the \textit{partie finie} (Pf) of the subsequent integrals (for details on taking the Pf, see \cite{Hadamard:1923, Damour:1988mr, Damour:2019lcq}). Prior to this, it is useful to introduce the inverse radial isotropic coordinate, $\bu = \br^{-1}$. Following these steps, one derives
\begin{align} \label{Parte Finie Map}
\begin{split}
    \theta_{n\mathrm{PM}}&(\gamma, \ell) = -\pi + \\ &2\sum_{k=0}^n \frac{1}{\ell^k} \binom{-\frac{1}{2}}{k} \mathrm{Pf}  \int_0^{p_\infty} \mathrm{d}\bu (p_\infty^2 - \bu^2)^{-\frac{1}{2} - k} \Tilde{w}(\gamma, \bu)^k,
\end{split}
\end{align}
where
\begin{align}
    \Tilde{w}(\gamma, \bu) = \sum_{i=1}^n \ell^{i-1} w_i (\gamma) \bu^i .
\end{align}
Upon matching terms at the same PM order, one finds that $w_i(\gamma)$ is completely determined by $\theta_k(\gamma) \,\,\, \forall \, k \in \{1,2,...,i\} \subset \mathbb{Z}$ (and vice versa). The mapping between these coefficients up to 6PM can be found in App.~\ref{app:pm_to_resummed_map}. 

With the calculation of the $w_i(\gamma)$ coefficients, we can now write the scattering angles in the $w^\mathrm{eob}$ model as
\begin{align}
    \label{Eq: eob scatter angle r}
    \theta_{n\mathrm{PM}}^{w^\mathrm{eob}}(\gamma, \ell) &= -\pi + 2\ell \int_{\brmin}^{\infty} \frac{\mathrm{d} \br}{\bar{r}^2 \sqrt{p_\infty^2 + w^\mathrm{eob}_{n\mathrm{PM}}(\gamma, \bar{r}) - \frac{\ell^2}{\bar{r}^2}}}.
\end{align}
In terms of $\bu$, this reads
\begin{align}
\label{Eq: weob resum scattering angles}
    \theta_{n\mathrm{PM}}^{w^\mathrm{eob}}(\gamma, \ell) &= -\pi + 2\ell \displaystyle\int_{0}^{\bar{u}_\mathrm{max}} \frac{\mathrm{d} \bu}{\sqrt{p_\infty^2 + w^\mathrm{eob}_{n\mathrm{PM}}(\gamma, \bu) - \ell^2 \bar{u}^2}},
\end{align}
where $\bar{u}_\mathrm{max}(\gamma, \ell) = 1/\bar{r}_\mathrm{min}(\gamma, \ell)$. 
In general, such expressions are highly non-trivial. In \cite{Damour:2022ybd}, several closed-form solutions were found using various substitution methods. Where possible, we use analytical solutions and revert to evaluating the scattering angles numerically when no such solution exists, as in \cite{Buonanno:2024vkx}.

\subsubsection{Reconstructing the NR Potentials} \label{Sec. Constructing NR Potentials}
We can use our sequence of NR simulations to extract a radiation-reacted gravitational potential that is central to understanding scattering \cite{Damour:2022ybd}.
The starting point follows the mass-shell condition, allowing us to re-express the radial momentum (Eq.~\ref{Eq: radial action}) as 
\begin{align}
    p_{\br}^2 &= p_\infty^2 - V_\mathrm{eff}(\br, \gamma, \ell),
\end{align}
where $V_{\rm eff}$ denotes an effective potential of the form
\begin{align}
V_\mathrm{eff}(\br, \gamma, \ell) &= \frac{\ell^2}{\br^2} - w(\br, \gamma).
\end{align}
This potential contains a term that looks like a centrifugal potential $\propto \ell^2 / \br^2$, encapsulating the complete dependence on $\ell$, and an energy-dependent radial potential $w(\br, \gamma)$ \cite{Damour:2022ybd}.

By inverting the relation between the radial potential and the scattering angle, i.e. Eq. \ref{Eq: weob Scattering Angle}, one obtains an implicit formula for the radial potentials in terms of the scattering angle
\begin{align} \label{Eq: Firsov's Inversion Formula}
    \nonumber
    w(\br,\gamma) &= p_\infty^2 \Bigg[ -1 \\
    &\quad + \exp\left( \frac{2}{\pi} \int_{\br\lvert p(\br,\gamma) \rvert}^\infty \mathrm{d}\ell \frac{\theta(\gamma, \ell)}{\sqrt{\ell^2 - \br^2 p^2(\br,\gamma)}} \right) \Bigg],
\end{align}
where $p(\br, \gamma) = \sqrt{p_\infty^2 + w(\br, \gamma)}$. This is an EOB analogue of Firsov's inversion formula \cite{Damour:2022ybd} (c.f. the original formula of Firsov \cite{Firsov:1953,Landau:1960mec}).

To determine $w(\br, \gamma)$ from our NR data, we follow the method introduced in \cite{Damour:2022ybd} and fit a template of the form 
\begin{align}
    \label{eq:L_resummed_template crit l}
     &\theta_{6\mathrm{PM},\mathrm{NR}}^\mathcal{L}(\gamma, \ell; \ell_{0,\mathrm{NR}}, \theta_{5, \mathrm{NR}}, \theta_{6, \mathrm{NR}}) 
     = \mathcal{L}\left(\frac{\ell_{0, \mathrm{NR}}}{\ell}\right) \times \notag
     \\ \nonumber
     &\quad \Bigg[ \hat{\theta}_{4\mathrm{PM}}(\gamma, \ell; \ell_{0, \mathrm{NR}})
     + 2\frac{\hat{\theta}_{5}(\gamma; \ell_{0, \mathrm{NR}}, \theta_{5, \mathrm{NR}})}{\ell^5}
     \nonumber \\
     &\qquad \qquad + 2\frac{\hat{\theta}_{6}(\gamma; \ell_{0, \mathrm{NR}}, \theta_{5, \mathrm{NR}}, \theta_{6, \mathrm{NR}})}{\ell^6} \Bigg],
\end{align}
where
\begin{align}
     \notag
     &\hat{\theta}_{5}(\gamma; \ell_{0, \mathrm{NR}}, \theta_{5, \mathrm{NR}}) = \theta_{5, \mathrm{NR}} - \frac{\ell_{0, \mathrm{NR}}}{2}\theta_4(\gamma) - \frac{\ell_{0, \mathrm{NR}}^2}{12}\theta_3(\gamma) \\ &\qquad \qquad - \frac{\ell_{0, \mathrm{NR}}^3}{24}\theta_2(\gamma) - \frac{19\ell_{0, \mathrm{NR}}^4}{720}\theta_1(\gamma),
\end{align}
and
\begin{align}
     \notag
     &\hat{\theta}_{6} (\gamma; \ell_{0, \mathrm{NR}}, \theta_{5, \mathrm{NR}}, \theta_{6, \mathrm{NR}}) = \theta_{6, \mathrm{NR}}  - \frac{\ell_{0,\mathrm{NR}}}{2}\theta_{5, \mathrm{NR}} 
     \\ &\qquad \quad - \frac{\ell_{0,\mathrm{NR}}^2}{12}\theta_4(\gamma) - \frac{\ell_{0,\mathrm{NR}}^3}{24}\theta_3(\gamma) \notag
     \\ &\qquad \qquad \quad - \frac{19\ell_{0,\mathrm{NR}}^4}{720}\theta_2(\gamma) - \frac{3\ell_{0,\mathrm{NR}}^5}{160}\theta_1(\gamma),
\end{align}
to our NR scattering angles. 
Calibrating the 3 parameters, ($\ell_{0,\mathrm{NR}}$, $\theta_{5, \mathrm{NR}}$ and $\theta_{6, \mathrm{NR}}$), we obtain a highly accurate fit to the scattering angles that can be used to reconstruct the NR radial potential, $w_{\rm NR} (\br, \gamma)$\footnote{We use the parametric representation introduced in \cite{Damour:2022ybd} that relates $w$ and $r$ to the Abel transformation of the scattering angles.}, via Eq. \ref{Eq: Firsov's Inversion Formula}.

In the following section, the validity of $w_{n\mathrm{PM}}^\mathrm{eob}$ is tested by comparison to $w_{\rm NR}$ (see Fig. \ref{Fig. EOB Potential Plots}). Additionally, we calibrate the $w^\mathrm{eob}$ model to NR data at 5PM via the EOB-PM potential (see, for example, Fig. \ref{Fig: weob Potential Fit Analysis}).

\subsubsection{Performance of the $w^\mathrm{eob}$ Resummation}

\begin{figure*}[htp]
  \centering
  \subfloat{
      \includegraphics[width=.5\textwidth]{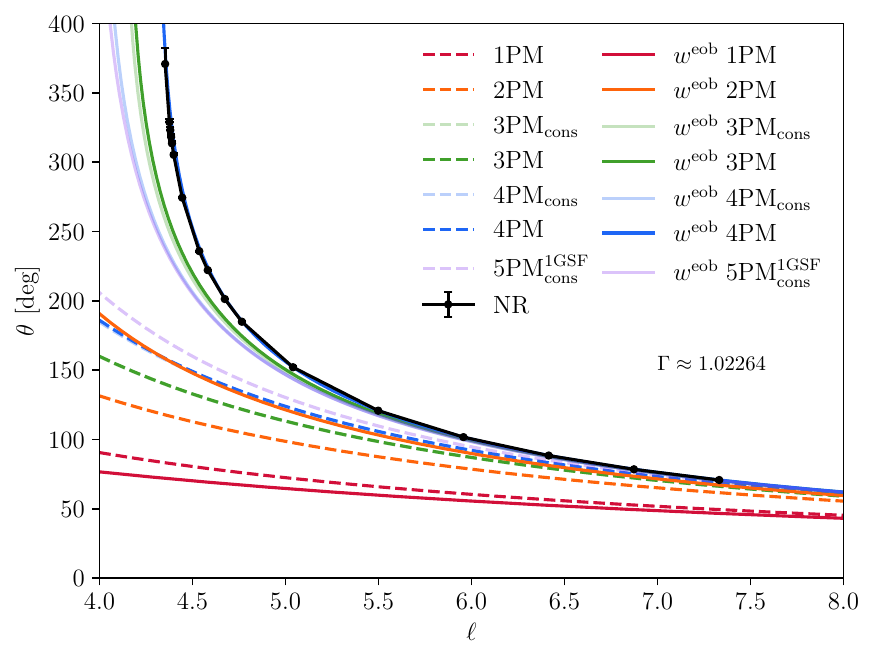}
  }
  \subfloat{
      \includegraphics[width=.5\textwidth]{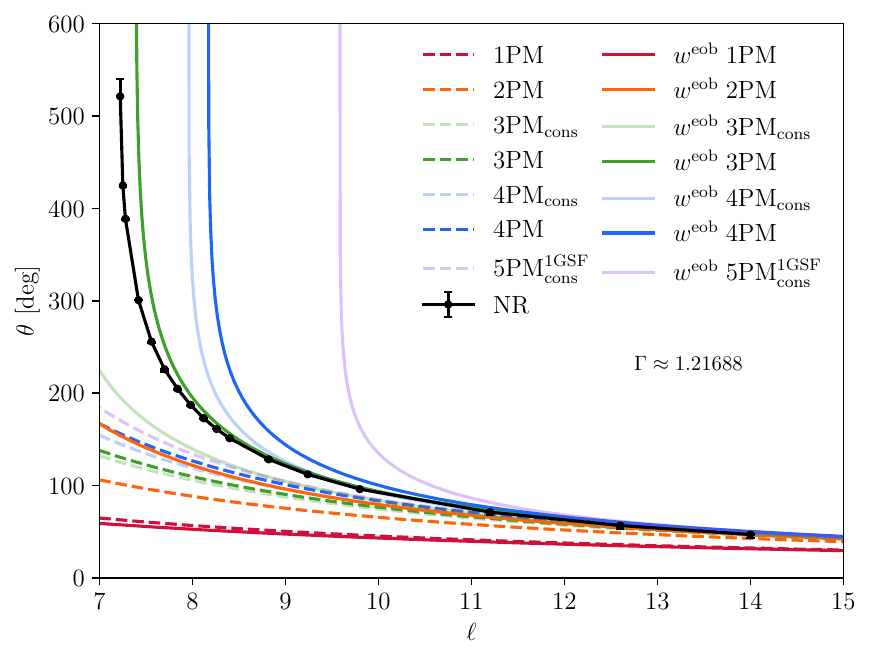}
  }    
  \caption{Comparison of the scattering angle predictions from the $w^\mathrm{eob}$ model with the scattering angles extracted from NR data. Left: Lowest energy, $\Gamma_1$, comparison. Right: Highest energy, $\Gamma_7$, comparison. We supplement both comparisons with the predictions of the PM-expanded scattering angles.}
  \label{Fig: weob Scattering Angles}
\end{figure*}

The work of \cite{Damour:2022ybd} showed that at the lowest energy, $\Gamma_1$, the $w^\mathrm{eob}$ model significantly improves on the performance of the PM-expanded scattering angles, with the 4PM predictions giving remarkably accurate predictions when compared to NR data (see Fig. \ref{Fig: weob Scattering Angles}). When analysed at the higher energies $\Gamma_2$ and $\Gamma_3$ (see Fig. \ref{Fig: Additional weob_plots}), 
the over-attractive nature of the 4PM term and the overly-repulsive nature of the 3PM term can become problematic~\cite{Rettegno:2023ghr}.
With partial access to the 5PM coefficients, and new NR data, we explore the $w^\mathrm{eob}$ model up to an energy of $\Gamma_7 \sim 1.21688$. 
The key results are shown in Fig.~\ref{Fig: weob Scattering Angles} as well as the supplementary plots in App.~\ref{Appendix: Additional weob_plots}.

\begin{figure}
    \centering
    \includegraphics[width=1\linewidth]{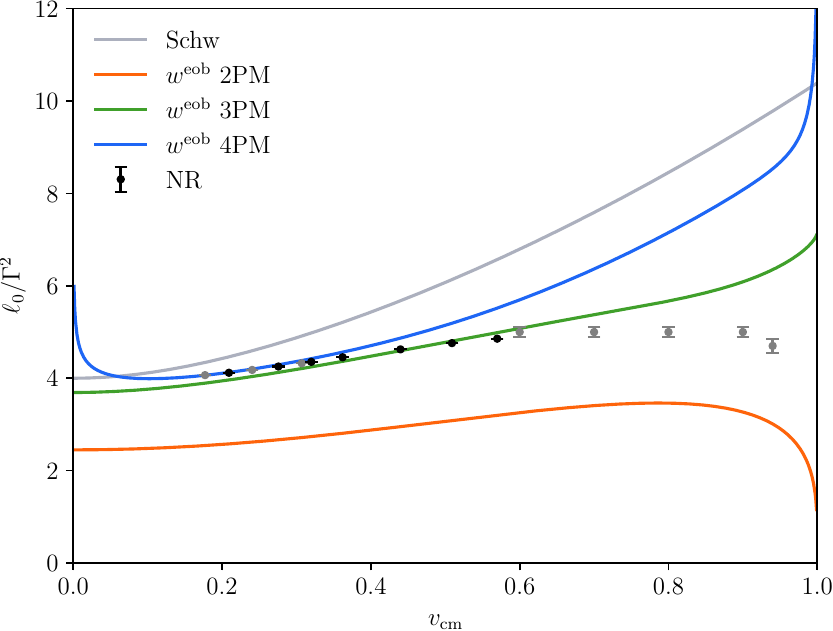}
    \caption{Re-scaled critical angular momentum predictions as a function of centre-of-mass velocity in the $w^\mathrm{eob}$ model. Here, we only plot the values for $\ell_0$ for PM orders at which we have a closed-form solution. As in Sec. \ref{Section: L-Resummed Comparison}, we overlay data points from \cite{Albanesi:2024xus, Shibata:2008rq, Sperhake:2009jz} in grey.}
    \label{fig:weob_critical_angular_momentum_plot}
\end{figure}

In Fig.~\ref{Fig: weob Scattering Angles} and Fig.~\ref{Fig: Additional weob_plots}, we show the behaviour of the $w^\mathrm{eob}$ model across all of our energies. 
Up to $\sim \Gamma_6$, the 4PM predictions transition from highly accurate to highly over-attractive, meanwhile the 3PM predictions transition from being under-attractive to accurate. 
At our highest energy, both the 3PM and 4PM predictions are over-attractive, a feature which is expected to be present beyond this energy. 
One might hope that the added information of the partial 5PM term might cure these issues. We see promising results between $\Gamma_1$-$\Gamma_4$, with the partial 5PM information behaving similarly to the conservative 4PM (3PM) predictions for $\Gamma_1$ ($\Gamma_2$). 
With 2GSF and radiative contributions required to complete the 5PM level, it is possible that a full 5PM term could give better agreement. 
Pushing to higher energies, we see that the partial 5PM term rapidly becomes over-attractive, with predictions at our highest energy heavily disagreeing with most of our NR data points. 

At low energies, the partial 5PM predictions exhibit seemingly unphysical behavior. 
Specifically, between $~ \gamma \sim 1.10364$ and $~ \gamma \sim 1.11364$, there exists a critical angular momentum, $\ell_0$. 
However, for $\ell < \ell_0$, the model does not predict a plunge and emulates a repulsive core (c.f. the 4PM conservative issues discussed in \cite{Damour:2022ybd}). 
Furthermore, below $~ \gamma \sim 1.10364$, the repulsive features dominate at small radii and no critical angular momentum is predicted, with $\theta^{w^\mathrm{eob}, \mathrm{1GSF}}_{5\mathrm{PM}, \mathrm{cons}} \rightarrow -\pi$ in the limit $\ell \rightarrow 0$. These undesirable features of $\theta^{w^\mathrm{eob}, \mathrm{1GSF}}_{5\mathrm{PM}, \mathrm{cons}}$ can likely be attributed to the divergent properties of $\theta_{5, \mathrm{cons}}^\mathrm{1GSF}$ in the low and high energy limits.

\begin{figure} 
  \centering
  \includegraphics[width=1\linewidth]{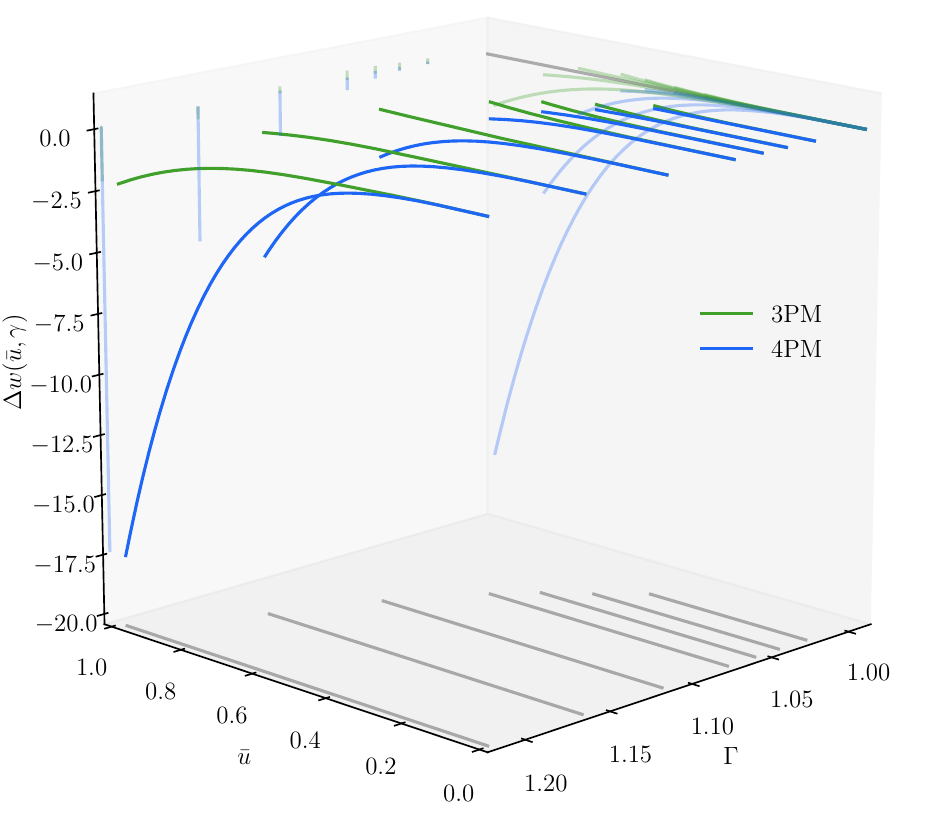}   
  \caption{Comparison of the EOB potential with NR data. EOB potential difference, $\Delta w(\bu, \gamma) = w_{\mathrm{NR}}(\bu, \gamma) - w_{n\mathrm{PM}}(\bu, \gamma)$, is shown as a function of the inverse isotropic radial coordinate for energies $\Gamma_1$ - $\Gamma_7$. We include 2D projections of the data onto each panel for clarity.
  }
  \label{Fig. EOB Potential Plots}
\end{figure}

As in Sec. \ref{Section: L-Resummed Comparison}, it is useful to analyse the critical angular momentum across our range of energies. In comparison to the $\mathcal{L}$-resummed scheme of Sec. \ref{Section: L-Resummed Comparison}, the $w^\mathrm{eob}$ model appears to be more robust across all energies at both 3PM and 4PM. Similarly, however, the $w^\mathrm{eob}$ model does not recover the apparent limit for $\ell_0$ as $v_\mathrm{cm} \rightarrow 1$ (as also discussed in \cite{Damour:2022ybd}). From Fig. \ref{fig:weob_critical_angular_momentum_plot}, it is expected that the 3PM and 4PM potentials will continue to become increasingly over-attractive in the high-velocity limit, while the accuracy in the low-velocity limit is not entirely clear and warrants further investigation. Similarly, we can summarise the physics of the $w^\mathrm{eob}$ model up to 4PM by comparing the EOB-PM potential against NR data. In Fig. \ref{Fig. EOB Potential Plots}, we see $w_{4\mathrm{PM}}$ agrees very well with $w_\mathrm{NR}$ at $\Gamma_1$, but very quickly shows over-attractive features compared to $w_\mathrm{NR}$. Meanwhile, $w_{3\mathrm{PM}}$ transitions from over-repulsive to over-attractive as we increase the energy, with the transition occurring near $\sim \Gamma_4$.

\begin{figure}[htp]
    \centering
    \includegraphics[width=1\linewidth]{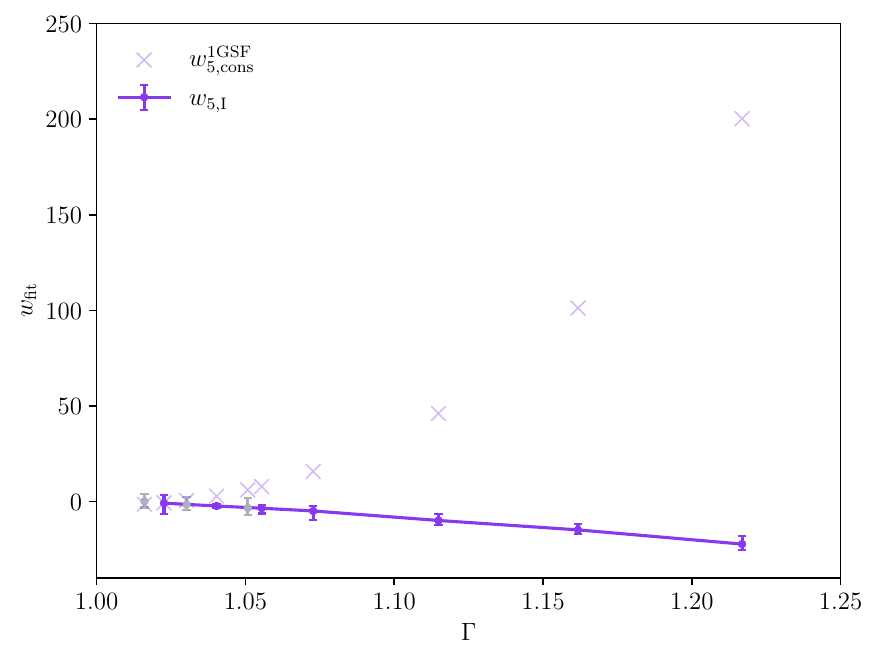}
    \caption{Values of $w_{5,\mathrm{I}}$ at energies $\Gamma_1$-$\Gamma_7$, compared with the values of $w_{5, \mathrm{cons}}^{1\mathrm{GSF}}$. Fits performed against data from \cite{Albanesi:2024xus} are overlaid in grey.}
    \label{fig:weob_fit_value}
\end{figure}

In a similar fashion to Sec. \ref{Section: L-Resummed Comparison}, we would like to understand the effects of higher order information on the $w^\mathrm{eob}$ model. We seek the ideal value of the $w_5(\gamma)$ coefficient by performing a fit to our NR data. It is non-trivial to use the scattering angles here, since the upper limit of integration depends on the value of $w_5(\gamma)$. Instead, we consider a template of the form
\begin{align} \label{Eq. weob fitting template}
    w_{5\mathrm{PM}, \mathrm{I}}(\bu, \gamma) = \sum_{i=1}^4 w_i(\gamma)\bu^i + w_{5,\mathrm{I}}\bu^5,
\end{align}
which is fitted against the NR extracted potentials, $w_\mathrm{NR}$.
In Fig. \ref{fig:weob_fit_value}, we show the recovered values of $w_{5,\mathrm{I}}$ against the values of $w_{5, \mathrm{cons}}^{1\mathrm{GSF}}$, with large discrepancies seen at high energies. 
Here, the errors of $w_{5, \mathrm{I}}$ arise due to applying the errors on $\theta_{6\mathrm{PM}}^\mathcal{L}$ to Eq. \ref{Eq: Firsov's Inversion Formula} \cite{Rettegno:2023ghr}. 
In Fig.~\ref{Fig: weob Potential Fit Analysis}, we calculate the EOB-PM potentials by inserting the recovered values of $w_{5, \mathrm{I}}$ into Eq. ~\ref{Eq. weob fitting template}, while in Fig.~\ref{Fig: weob Scattering Angle Fit Analysis} we calculate the scattering angles by inserting these fitted EOB-PM potentials into Eq.~\ref{Eq: eob scatter angle r} (we denote these scattering angles $\theta^{w^\mathrm{eob}}_{5\mathrm{PM, I}}$).  

Comparing these models against the NR data, we find that at $\Gamma_1$, the value of $w_{5, \mathrm{I}}$ is negligible, since the 4PM predictions are already highly accurate. 
In contrast, the fitted value of $w_{5, \mathrm{I}}$ at $\Gamma_7$ leads to repulsive attributes. 
This is likely due to performing our fit using Eq. \ref{Eq. weob fitting template}, {with divergent behaviour in the $\br \rightarrow 0$ being dictated by $w_{5,\mathrm{I}}$}\cite{Rettegno:2023ghr}. 
In particular, we see discrepancies with NR data before the plunge, with the predictions of $\theta^{w^\mathrm{eob}}_{5\mathrm{PM, I}}$ being over-attractive up to $\ell \sim 7.6$. 

As discussed in Sec. \ref{Sec. Performance of the L Resummation}, the physical interpretation of the pseudo-PM terms must be approached with caution, since calibration to NR causes non-perturbative information to propagate into such terms.

\subsubsection{Pad\'{e} resummation of the $w$-potentials}

As discussed in \cite{Rettegno:2023ghr}, we have been using PM-expanded EOB radial potentials of the form $w(\bu) \sim \bu + \bu^2 + \cdots$. However, we could further resum the EOB-PM potentials to help mitigate against the slow-convergence and improve the stability of the potentials towards the strong-field regime. 
Here, we explore the use of Pad\'e resummation as applied to the EOB-PM potentials and how this translates into the predicted scattering angles.

Following \cite{Damour:1997ub}, we let $w_n (\bu) = w_0 + w_1 \bu + \cdots + w_n \bu^n$ denote the PM (Taylor) series truncated at $\mathcal{O}(G^n)$. A Pad\'e approximant of order $m + k = n$ can be defined such that 
\begin{align}
P^m_k (\bu) &= \frac{N_m (\bu)}{D_k (\bu)}, 
\end{align}
subject to the constraint that $T_n [ P^m_k (\bu) ] = w_n (\bu)$, where $T_n [ \cdots ]$ denotes the Taylor expansion operator. 
When calibrating the potentials against NR, there are a few different approaches we could follow. 
For example, we may choose to Padé resum Eq. \ref{Eq. weob fitting template} \textit{after} calibrating the 5PM term in the EOB-PM potential, which we will denote \( w_{5\mathrm{PM}, \mathrm{II}} \).
Alternatively, we could apply Padé resummation to Eq. \ref{Eq. weob fitting template} and calibrate the free 5PM coefficient that appears in the resulting approximant, which will be denoted by $w_{5\mathrm{PM}, \mathrm{III}}$. For both $w_{5\mathrm{PM}, \mathrm{II}}$ and $w_{5\mathrm{PM}, \mathrm{III}}$, we find that a $P^4_1$ Pad\'e approximant yields the best results, though an exhaustive exploration across the parameter space is beyond the scope of this paper. 

\begin{figure*}[htp]
  \centering
  \subfloat{
      \includegraphics[width=.5\textwidth]{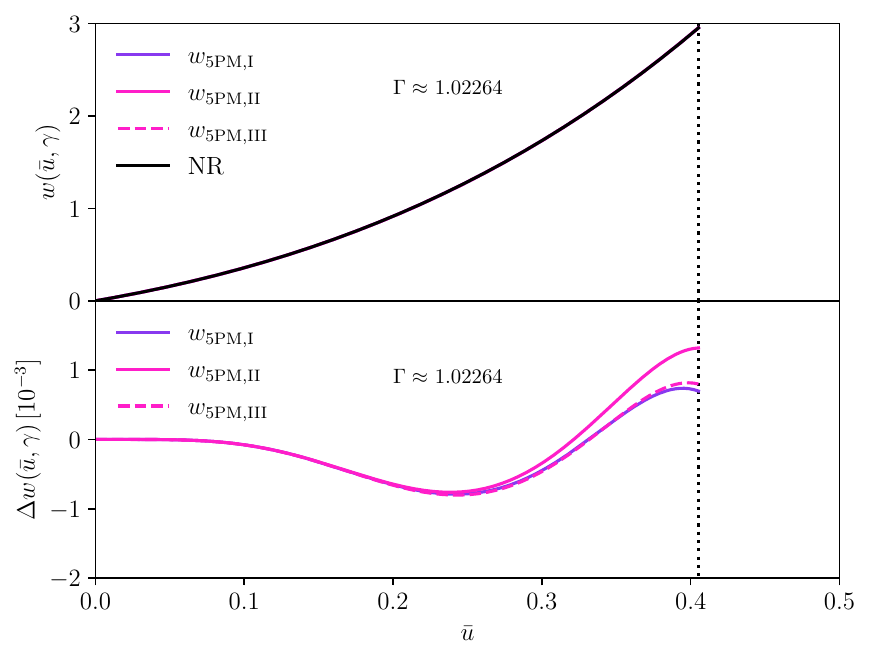}
  }
  \subfloat{
      \includegraphics[width=.5\textwidth]{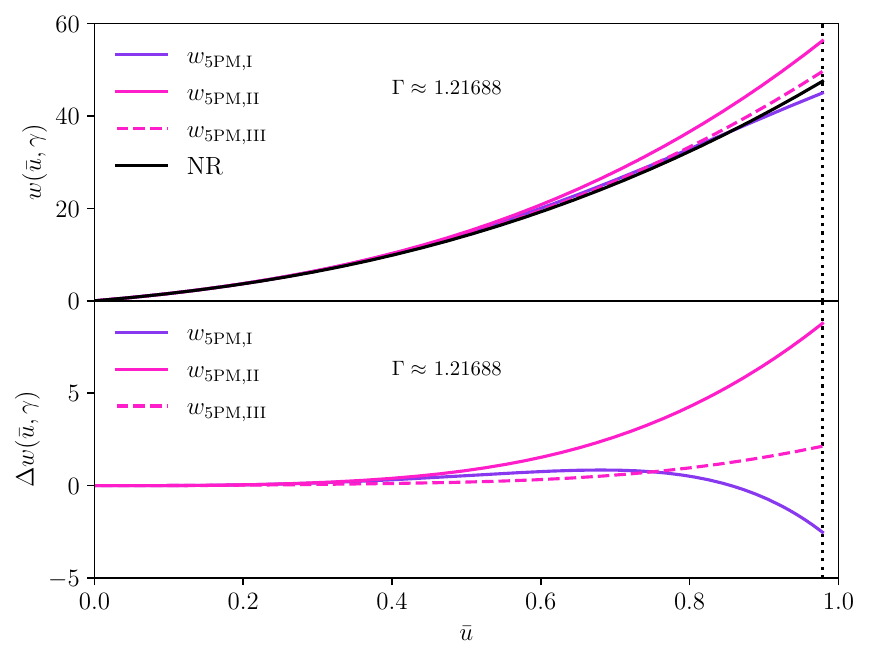}
  }    
  \caption{Performance of the $w^\mathrm{eob}$ 5PM fitting procedures, $w_{5\mathrm{PM}, \mathrm{I}}$, $w_{5\mathrm{PM}, \mathrm{II}}$ and $w_{5\mathrm{PM}, \mathrm{III}}$ against NR potentials (obtained via the process outlined in Sec. \ref{Sec. Constructing NR Potentials}). Left: Comparison at lowest energy, $\Gamma_1$. Right: Comparison at highest energy, $\Gamma_7$. The vertical dotted line denotes the value of $\bu$ up to which our NR data is valid.}
  \label{Fig: weob Potential Fit Analysis}
\end{figure*}

In Fig. \ref{Fig: weob Potential Fit Analysis}, we see that at the lowest energy the inclusion of a 5PM term, and the exact prescription for resumming the EOB-PM potentials, has little impact on the overall agreement with NR.
However, at $\Gamma_7$, we start to see some noticeable differences. 
First, we had already seen that the standard EOB-PM potential had an unphysical turning point. 
Second, the expression derived by calibrating the 5PM term and then taking the Pad\'e approximant, $w_{5\mathrm{PM}, \mathrm{II}}$, is giving seemingly worse results than the expression derived by calibrating the 5PM term in the Pad\'e approximant itself, $w_{5\mathrm{PM}, \mathrm{III}}$, \textit{and} the standard EOB-PM potential. 
Nonetheless, the Pad\'e approximant shows significantly improved agreement against NR towards small radii. 

\begin{figure*}[htp]
  \centering
  \subfloat{
      \includegraphics[width=.5\textwidth]{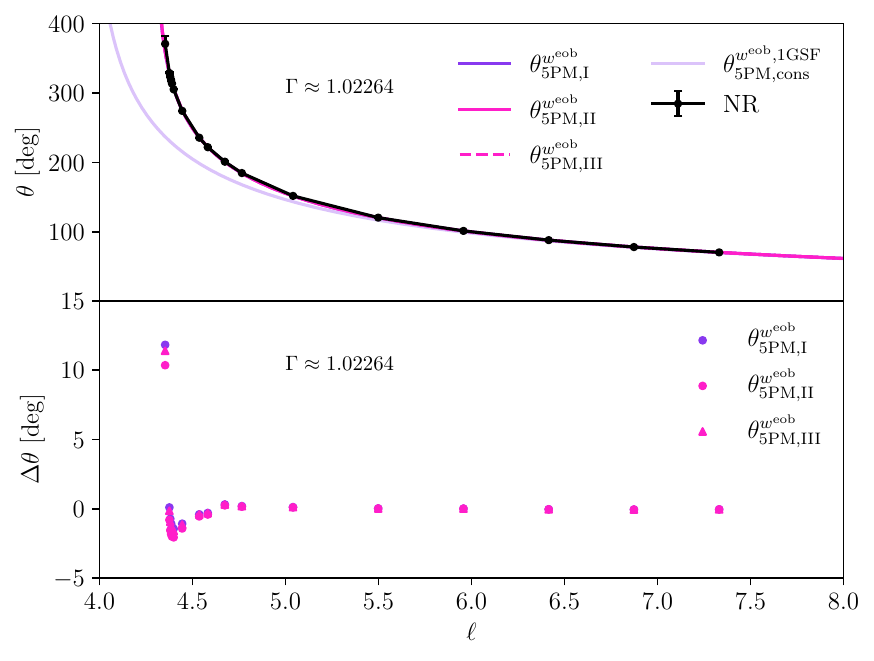}
  }
  \subfloat{
      \includegraphics[width=.5\textwidth]{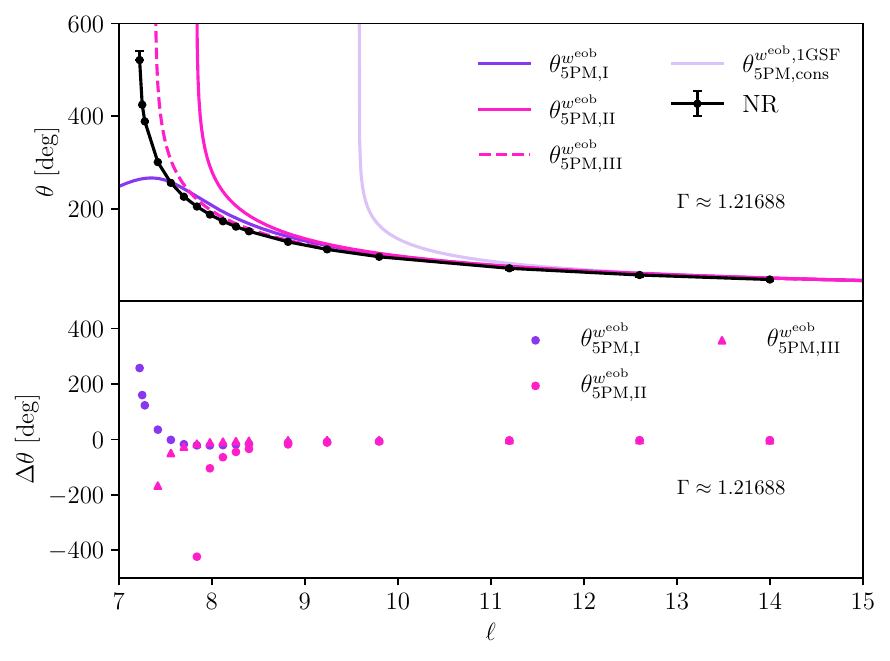}
  }    
  \caption{Analysing the effects of $w_{5\mathrm{PM}, \mathrm{I}}$, $w_{5\mathrm{PM}, \mathrm{II}}$ and $w_{5\mathrm{PM}, \mathrm{III}}$ on the scattering angle (Eq. \ref{Eq: weob Pade resum scattering angles}). For comparison, we plot these alongside $\theta_{5\mathrm{PM},\mathrm{cons}}^{w^\mathrm{eob}, 1\mathrm{GSF}}$.}
  \label{Fig: weob Scattering Angle Fit Analysis}
\end{figure*}

To investigate how these potentials alter the scattering angles, we again evaluate the scattering integral
\begin{align}
\label{Eq: weob Pade resum scattering angles}
    &\theta_{5\mathrm{PM},X}^{w^\mathrm{eob}}(\gamma, \ell) = -\pi
    \nonumber  \\
    &\quad + 2\ell \displaystyle\int_{0}^{\bar{u}_\mathrm{max, X}} \frac{d \bu}{\sqrt{p_\infty^2 + w^\mathrm{eob}_{5\mathrm{PM}, \mathrm{X}}(\gamma, \bu) - \ell^2 \bar{u}^2}},
\end{align}
where $\mathrm{X} \in \{\mathrm{II},\mathrm{III}\}$ denotes the type of fit considered and $\bu_\mathrm{max, \mathrm{X}}$ is defined as the smallest, real, positive root of $p_\infty^2 + w^\mathrm{eob}_{5\mathrm{PM}, \mathrm{X}}(\gamma, \bu) - \ell^2 \bar{u}^2$. 
The resulting scattering angles are shown in Fig.~\ref{Fig: weob Scattering Angle Fit Analysis}.
In particular, we again see that the scattering angles are in broad agreement with NR at the lowest energy, as expected. 
How we choose to include and calibrate the 5PM term has little impact.

Based on the discussion above, we naturally anticipate larger discrepancies occuring at the highest energies. The NR-tuned Padé approximant shows the best performance, both in terms of its agreement against the NR scattering angles as well as avoiding the unphysical turnover predicted by the EOB-PM scattering angles. 
Whilst the Pad\'e approximant derived from the NR-tuned EOB-PM potentials shows the worst agreement, it does not exhibit an unphysical repulsive core, suggesting a more robust model.  
We regard these results as a proof-of-principle that Pad\'e resummation of the EOB-PM potentials is a useful strategy for further improving the accuracy and robustness of the PM-expanded potentials. 
We leave a more detailed discussion to future work.

\subsection{The SEOB-PM Model}

\subsubsection{The SEOB-PM Resummation of the Scattering Angle}

An EOB Hamiltonian based on the PM expansion was recently presented in ~\cite{Buonanno:2024vkx}, incorporating full non-spinning and spinning information at (physical) 4PM and partial information at 5PM. The SEOB-PM Hamiltonian is based on a \textit{test-mass} in a deformed \textit{Kerr} spacetime ~\cite{Damour:2001tu,Damour:2014sva,Balmelli:2015zsa,Damour:2008qf,Nagar:2011fx,Khalil:2023kep,Pompili:2023tna,Ramos-Buades:2023ehm}. The PM information is incorporated using the post-Schwarzschild$\ast$ (PS$\ast$) gauge, in which the non-geodesic term $\hat{Q}(r,\gamma)$ is set to zero. In the non-spinning limit, i.e. $a_{\pm} \rightarrow 0$, the Hamiltonian can be expressed in terms of the metric potentials $A(r)$ and $B(r)$ \cite{Buonanno:2024vkx}
\begin{align}
    H_{\rm eff} &= \sqrt{A(r) \left( 1 + \frac{\ell^2}{r^2} + \frac{p^2_r}{B(r)} \right)},
\end{align}
where, in the PS$\ast$ gauge, $A(r)$ and $B(r)$ are the canonical Schwarzschild metric potentials.
As $H_{\rm eff} = E_{\rm eff}$, the equation can be inverted to solve for $p_r$ \cite{Buonanno:2024vkx}
\begin{align}
    p^2_r &= \frac{1}{B(r)} \Bigg[ \frac{\gamma^2}{A(r)} - \left( 1 + \frac{\ell^2}{r^2}\right) \Bigg],
    \\
    &= p^2_{\infty} - \frac{\ell^2}{r^2} + w_{\rm{SEOB-PM}} \left(\gamma, \ell, r\right),
\end{align}
\textit{de facto} defining a new resummed effective potential $w_{\rm SEOB-PM}$. The PM information is incorporated through PM expansions of the $A$-potential, i.e. $A(r) \rightarrow A(r) + \Delta A$, where
\begin{align}
    \Delta A = \sum_{{i = 2}} \frac{A_i(\gamma)}{r^i},
\end{align}
such that the test-mass limit of the model is recovered in the limit $\Delta A \rightarrow 0$. As in Sec. \ref{SubSec. weob}, this results in
\begin{align}
    w_{\rm{SEOB-PM}} (\gamma, \ell, r) = \sum_{i} \frac{w^\ast_i(\gamma)}{r^i},
\end{align}
and
\begin{align}
    w_{\rm{SEOB-PM}}^{n\mathrm{PM}} (\gamma, \ell, r) = \sum_{i=1}^n \frac{w^\ast_i(\gamma)}{r^i}.
\end{align}
As before, we use Eq. \ref{Eq: Hamilton-Jacobi Scattering Angle} to obtain
\begin{align} \label{Eq. SEOB Scattering Angles}
\begin{split}
    \theta(\gamma, &\ell) = -\pi + \\ 
    &2\ell \int_{r_\mathrm{min} (\gamma, \ell)}^{\infty} \frac{\mathrm{d} r}{r^2 \sqrt{p_\infty^2 + w_\mathrm{SEOB-PM}(\gamma, r, \ell) - \frac{\ell^2}{r^2}}},
\end{split}
\end{align}
where $r_\mathrm{min} (\gamma, \ell)$ is defined as the largest real, positive root of $p_{r}$.
We recover the $w^\ast_i(\gamma)$ coefficients by mapping directly to the {$\theta_k(\gamma)$} coefficients using
\begin{align} 
\label{Parte Finie Map SEOB-PM}
\begin{split}
    \theta_{n\mathrm{PM}}&(\gamma, \ell) = -\pi + \\ &2\sum_{k=0}^n \frac{1}{\ell^k} \binom{-\frac{1}{2}}{k} \mathrm{Pf}  \int_0^{p_\infty} \mathrm{d}u (p_\infty^2 - u^2)^{-\frac{1}{2} - k} \Tilde{w}^\ast(\gamma, u)^k,
\end{split}
\end{align}
where
\begin{align}
    \Tilde{w}^\ast(\gamma, u) = \sum_{i=1}^n \ell^{i-1} w^\ast_i (\gamma) u^i .
\end{align}
Inserting $w^{n\mathrm{PM}}_\mathrm{SEOB-PM}$ into Eq. \ref{Eq. SEOB Scattering Angles} in place of {$w_\mathrm{SEOB-PM}$}, we get
\begin{align}
    &\theta_{n\mathrm{PM}}^\mathrm{SEOB-PM}(\gamma, \ell) = -\pi \nonumber \\
    &\quad + 2\ell \int_{r_\mathrm{min}}^{\infty} \frac{\mathrm{d} r}{r^2 \sqrt{p_\infty^2 + w^{n\mathrm{PM}}_\mathrm{SEOB-PM}(\gamma, r, \ell) - \frac{\ell^2}{r^2}}},
\end{align}
or in terms of the inverse radial coordinate $u$,
\begin{align}
\label{Eq: seob resum scattering angles}
    &\theta_{n\mathrm{PM}}^\mathrm{SEOB-PM}(\gamma, \ell) = -\pi \nonumber  \\
    &\quad + 2\ell \displaystyle\int_{0}^{u_\mathrm{max}} \frac{\mathrm{d} u}{\sqrt{p_\infty^2 + w^{n\mathrm{PM}}_\mathrm{SEOB-PM}(\gamma, u, \ell) - \ell^2 u^2}}.
\end{align}
Due to the structure of the potential, it is not trivial to find a closed-form analytical solution of the scattering angle. Instead, we follow \cite{Buonanno:2024vkx} and evaluate the integral numerically. 

\subsubsection{Performance of the SEOB-PM Resummation}

\begin{figure*}[htp]
  \centering
  \subfloat{
      \includegraphics[width=.5\textwidth]{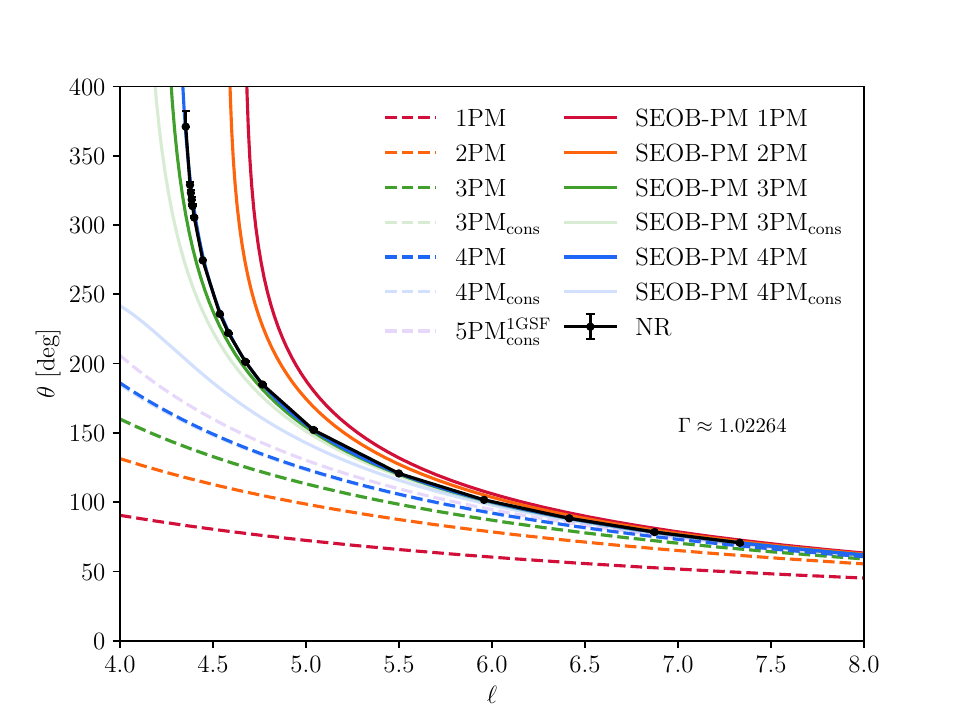}
  }
  \subfloat{
      \includegraphics[width=.5\textwidth]{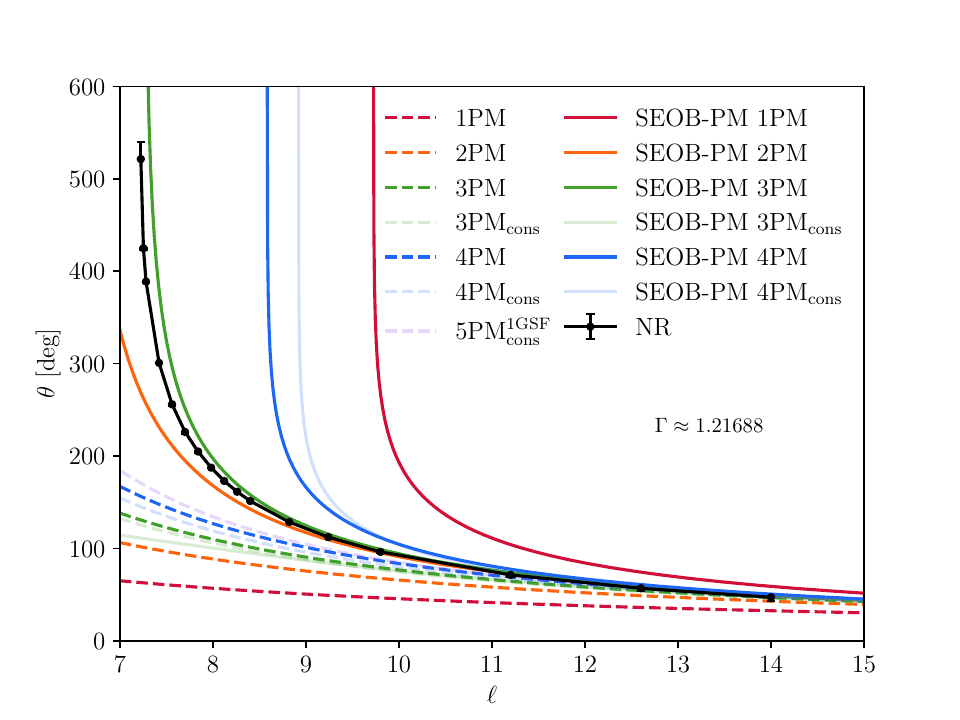}
  }    
  \caption{Comparison of the scattering angle predictions from the SEOB-PM model with the scattering angles extracted from NR data, supplemented with the PM-expanded scattering angles as in Fig. \ref{fig:L_resummed_vs_nr_plots}. Left: Comparison at lowest energy, $\Gamma_1$. Right: Comparison at highest energy, $\Gamma_7$.}
  \label{Fig: seob Scattering Angles}
\end{figure*}

\begin{figure*}[htp]
  \centering
  \subfloat{
      \includegraphics[width=.5\textwidth]{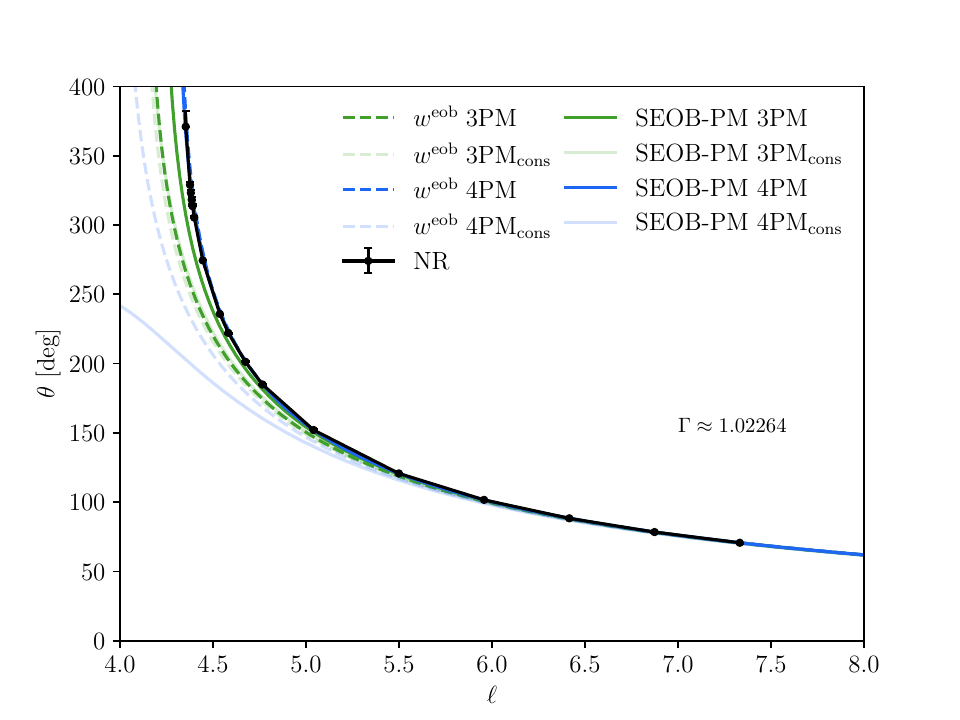}
  }
  \subfloat{
      \includegraphics[width=.5\textwidth]{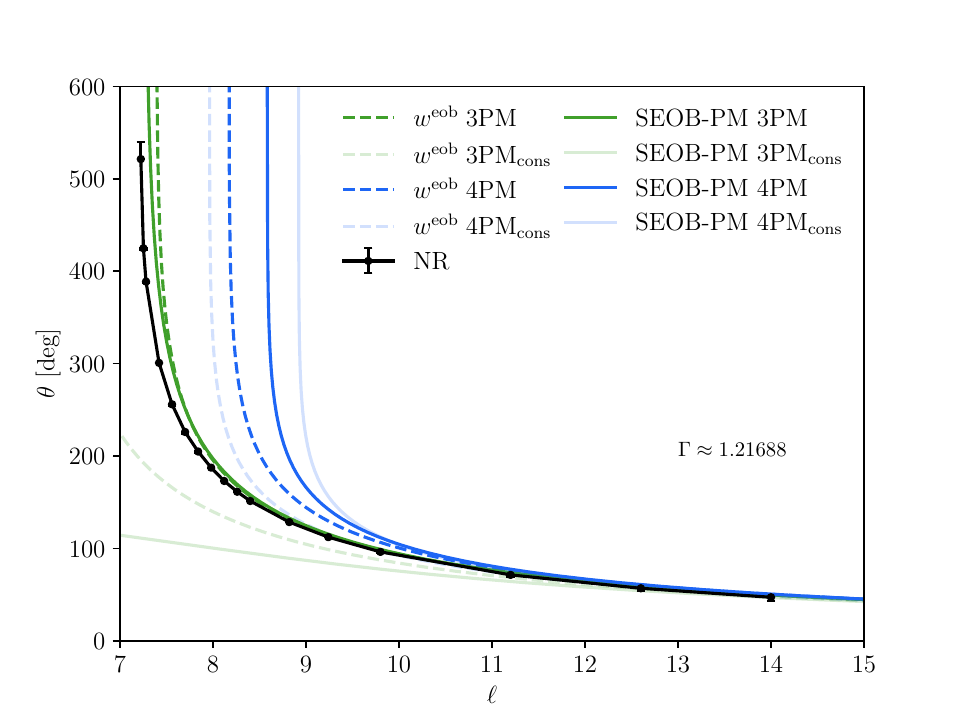}
  }    
  \caption{Direct comparison of the SEOB-PM model with the $w^\mathrm{eob}$ model at 3PM and 4PM, with NR data plotted for reference. Left: Comparison at lowest energy, $\Gamma_1$. Right: Comparison at highest energy, $\Gamma_7$.}
  \label{Fig: seob_v_weob Scattering Angles}
\end{figure*}

In \cite{Buonanno:2024vkx}, the complete non-spinning SEOB-PM model was presented at $\Gamma_1$ up to 4PM, with comparisons against the $w^\mathrm{eob}$ model at $\Gamma_2$ and $\Gamma_3$. 
In this section, we extend these results to show the performance of the SEOB-PM model against the available NR simulations, including a new comparison between the SEOB-PM model and the $w^\mathrm{eob}$ model at high energies. 
The key results are shown in Fig. \ref{Fig: seob Scattering Angles} and Fig. \ref{Fig: seob_v_weob Scattering Angles}, as well as the supplementary plots in App. \ref{Appendix: Additional seob_plots}.

The SEOB-PM model gives excellent agreement with NR data at low energies (up to $\sim \Gamma_4$), in particular at the 4PM level. 
Pushing to higher energies, the 4PM predictions start to become increasingly over-attractive while the 3PM results transition from over-attractive to over-repulsive. We find, as with the $w^\mathrm{eob}$ model, that the naively expected hierarchical structure across the PM orders does not hold at high energies.  
Interestingly, we find that the 2PM model outperforms the 4PM accurate model above $\sim \Gamma_5$, especially in the strong-field regime. 
As we have seen for the $w^\mathrm{eob}$ model, ignoring radiative contributions typically introduces unphysical features into the model. 
Likewise, the conservative sector of the 3PM and 4PM contributions does not accurately agree with the NR data, demonstrating the importance of including radiative corrections. 
Whereas at low energies, the 3PM and 4PM conservative predictions exhibit repulsive behaviour in the strong field regime. 
As we transition to higher energies, the 3PM conservative predictions become increasingly repulsive, while the 4PM conservative term transitions to highly over-attractive rapidly.

The SEOB-PM model differs from the $w^\mathrm{eob}$ model in two key ways. 
First, the SEOB-PM Hamiltonian reduces to that of a test-mass in Kerr spacetime in the probe limit. Second, the $w^\mathrm{eob}$ uses the post-Schwarzschild gauge and includes PM information via the non-geodesic $Q$ term, whereas the SEOB-PM model works in the post-Schwarzschild$\ast$ gauge and modifies the $A$ potential. 
Analysing Fig.~\ref{Fig: weob Scattering Angles} and Fig.~\ref{Fig: seob Scattering Angles} (and likewise Fig.~\ref{Fig: Additional weob_plots} and Fig.~\ref{Fig: Additional seob_plots}) highlights the significance of these differences at low PM orders. 
The 1PM and 2PM accurate SEOB-PM model shows a significant improvement in accuracy compared to the $w^\mathrm{eob}$ model across all energies. Interestingly, at 3PM and 4PM, the two models show remarkable similarities. 
In Fig. \ref{Fig: seob_v_weob Scattering Angles}, we directly compare the $w^\mathrm{eob}$ model with the SEOB-PM model at the 3PM and 4PM level, across our energy range. For brevity, we only show the results at $\Gamma_1$ and $\Gamma_7$. At our lowest energy, both models agree with NR data to a high degree of accuracy, with SEOB-PM showing minimal improvements over $w^\mathrm{eob}$. Up to $\sim \Gamma_3$, this behaviour largely remains intact, with the 3PM predictions of $w^\mathrm{eob}$ becoming more accurate than those of SEOB-PM only near $\Gamma_3$. 

Transitioning to higher energies, larger discrepancies appear, with the 3PM and 4PM levels of $w^\mathrm{eob}$ outperforming the corresponding predictions of SEOB-PM. At the highest energy, the relative performance of each model changes once again. 
Restricting to 3PM, the SEOB-PM model demonstrates stronger convergence to NR data than the $w^\mathrm{eob}$ model. 
In contrast, the divergent behavior at 4PM is seemingly stronger in the SEOB-PM model compared to $w^\mathrm{eob}$, leading to a loss in accuracy. 
The post-Schwarzschild gauge choice of the $w^\mathrm{eob}$ model appears to be more robust at 4PM, whilst providing comparable levels of accuracy to the SEOB-PM model.
However, we caveat that the SEOB-PM model enforces the correct test-particle limit in contrast to $w^\mathrm{eob}$ and a detailed examination at higher mass ratios would be of interest.

\section{Conclusions}

We have presented new high-energy NR simulations of non-spinning, equal-mass binary black hole scattering events, extending the available suite of simulations \cite{Damour:2014afa, Rettegno:2023ghr} up to a Lorentz factor of $\gamma = 1.96159$ ($\Gamma_7 = 1.21688$). 
Using these simulations, we comprehensively assessed three recently proposed resummations of the PM-expanded scattering angle: the $\mathcal{L}$-resummation \cite{Damour:2022ybd}, the $w^\mathrm{eob}$ model \cite{Damour:2022ybd} and the SEOB-PM model \cite{Buonanno:2024vkx}. 

The $\mathcal{L}$-resummation exploits the singular behaviour in the geodesic limit to improve the PM-expanded scattering angles. Two key ingredients are high-order PM information and a determination of the critical angular momentum that marks the transition from scattering orbits to bound orbits.
The framework demonstrates high accuracy at low energies, where the PM coefficients are well behaved. 
However, as is apparent from Fig. \ref{fig:L_resummed_critical_angular_momentum}, this scheme is not robust across all energies due to the change in PM hierarchy, which is related to the low- and high-energy behaviour of the PM coefficients. 
The inclusion of recently derived partial 5PM information improves accuracy at lower energies, but the subsequent Cauchy estimate of $\ell_0$ again diverges in the low- and high-velocity limits. 
It remains to be seen if radiative corrections to the partial 5PM term mitigates these divergences, which is not the case at 4PM. 

To investigate the effects of higher order PM information, we performed two fits to our NR data, following \cite{Damour:2022ybd}. 
In the first approach, we just explore the role of higher-order PM information in the $\mathcal{L}$-resummation scheme, estimating the critical angular momentum using the Cauchy estimate derived from the PM coefficients and an NR calibrated psuedo-5PM term, $\theta_{5,\mathrm{I}}$. 
The second approach takes a more agnostic viewpoint, where both the pseudo-5PM coefficient and the critical angular momentum are taken to be free coefficients that are calibrated against NR. 
We find that the Cauchy estimate of $\ell_0$ may be the dominating factor in the poor performance of the $\mathcal{L}$-resummation, with the NR calibrated critical angular momentum offering a significant improvement in accuracy.  
However, we again stress that while the fits presented give us some insight into the behavior of higher PM orders, they are effective parameters and we should treat their physical interpretation with caution. 

The $w^\mathrm{eob}$ scheme demonstrates high accuracy at low energies. As the energy increases, {undersirable} behavior emerges, particularly at the 4PM level. 
Compared to the $\mathcal{L}$-resummation, the $w^\mathrm{eob}$ model demonstrates more hierarchical stability, but the 4PM term does become increasingly over-attractive at high energies. 
Incorporating partial 5PM information has adverse effects at low energies. 
Repulsive features dominate in the strong-field regime and, at very low energies, exclude the possibility for a transition from unbound to bound orbits.  Between $\Gamma_1$-$\Gamma_4$, the partial 5PM contribution leads to more accurate predictions, but lacks stability as the energy is increased further. 
We can likely attribute this behavior to the power-law divergence present in the $\theta_{5, \mathrm{cons}}^\mathrm{1GSF}$ term. 

We outlined a procedure for incorporating NR information into the PM-expanded EOB potentials, recovering a pseudo-5PM coefficient that is tuned against NR.  
This potential was shown to produce an accurate series of scattering angles at low energies, but is still comparatively poor at higher energies. 
Motivated by \cite{Rettegno:2023ghr}, we explored the possibility of using Pad\'e resummation to improve the accuracy and stability of the EOB potentials. 
In particular we found that by performing a Pad\'e resummation of the PM-expanded EOB potential, before we tuned the free-coefficients against NR, led to a significant increase in the accuracy and stability of the model, even at the highest energy. 
A more thorough exploration of this procedure is left to future work. 

The SEOB-PM offers an alternative resummation of the scattering angle. While sharing many features with the $w^\mathrm{eob}$ model, the SEOB-PM model differs in two fundamental ways: Incorporation of PM information via the PS$\ast$ gauge, and (by construction) the ability to recover the test-mass dynamics. Like the $w^\mathrm{eob}$ model, the SEOB-PM model demonstrates a strong level of convergence with NR data at low energies, but contains {undesirable} features when tested at high energies. 
In general, the SEOB-PM hierarchy is less stable than that of $w^\mathrm{eob}$, with 
2PM and 3PM out-performing 4PM around $\sim \Gamma_{5/6}$. 
Furthermore, the conservative sector of the SEOB-PM model lacks stability compared to the $w^\mathrm{eob}$ across our parameter space, at both the 3PM and 4PM levels. 
Direct comparison with the $w^\mathrm{eob}$ shows the 4PM predictions are more robust in the PS gauge at higher energies, while the choice of PS over PS$\ast$ generates minimal impact at low energies. 

The work presented here can be expanded in several directions. 
Having demonstrated the significance of $\ell_0$ within the $\mathcal{L}$-resummation scheme, it would be beneficial to improve analytical estimates of the critical angular momentum, for instance by incorporating insights from the GSF framework \cite{Long:2024ltn}.
For the $w^\mathrm{eob}$ model, further investigation into the improvement of scattering angles associated with a Pad\'e-resummed potential is warranted.
In particular, the robustness of the scheme needs to be tested at all available PM orders and extended to spinning, unequal mass systems, which will be the focus of future work.
It would be interesting to explore the low-energy behaviour of the scattering angles in more detail. 
With complete 5PM results on the horizon, it will be important to explore the complete mass-ratio and spin dependence of the resummed PM angles and validate results against the non-perturbative information extracted from NR. 
Whilst a comparison against the scattering angles suggests that the 4PM information may be sufficient at lower energies, we need to understand how this impacts the accuracy of semi-analytical waveform models for bound-orbit compact binaries, e.g. \cite{Buonanno:2024byg}. 

Our suite of high-energy NR simulations indicate a clear need for higher-order PM information, highlighting the importance and value of of high-order analytical calculations in support of the development of fast and accurate semi-analytical waveform models.

\section*{Acknowledgements}
The authors thank Alessandra Buonanno for useful discussions. We also thank Gregorio Carullo, Christopher Whittall, Alice Bonino and Adam Clark for comments on the manuscript.
S.S. and G.P. acknowledge support from a Royal Society University Research Fellowship URF{\textbackslash}R1{\textbackslash}221500 and RF{\textbackslash}ERE{\textbackslash}221015. 
P.S. acknowledges support from a Royal Society Research Grant RG{\textbackslash}R1{\textbackslash}241327 and from STFC grant ST/V005677/1.
G.P. and P.S. acknowledge support from STFC grant ST/Y00423X/1.
Numerical simulations and computations were performed using the University of Birmingham's BlueBEAR HPC, which provides a High Performance Computing service to the University's research community, the Sulis Tier 2 HPC platform hosted by the Scientific Computing Research Technology Platform at the University of Warwick, funded by EPSRC Grant EP/T022108/1 and the HPC Midlands+ consortium, and on the Bondi HPC cluster at the Birmingham Institute for Gravitational Wave Astronomy.

\clearpage 

\appendix 

\section{Analysing Junk Radiation}
\label{Appendix: Junk}
\begin{figure}[htp]
    \centering
    \includegraphics[width=1\linewidth]{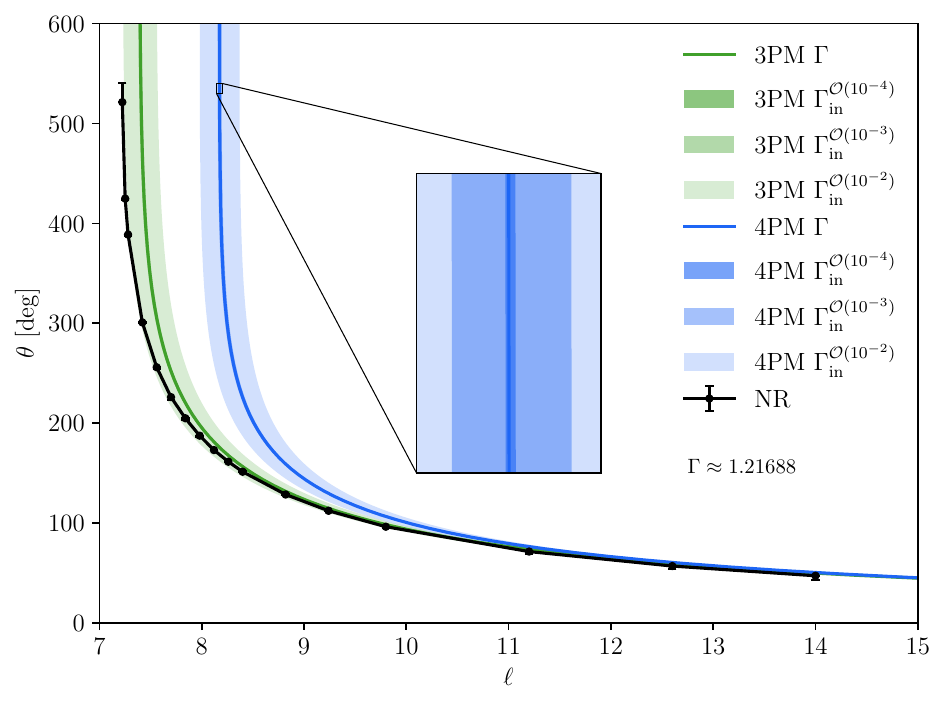}
    \caption{Effect of junk radiation on the scattering angle predictions at $\Gamma_7$. The notation $\Gamma_\mathrm{in}^{\mathcal{O}(10^{-N})}$ is shorthand for $\Gamma_\mathrm{in} \pm \mathcal{O}(10^{-N})$.}
    \label{fig:weob_junk_comparison}
\end{figure}

The initial data produced by the \texttt{TwoPunctures} thorn \cite{Ansorg:2004ds} are based on the Bowen-York decomposition of the constraint equations \cite{Bowen:1980yu, Brandt:1997tf}. This corresponds to a specific choice for the freely specifiable data in the conformal transverse-traceless decomposition of the Einstein field equations \cite{York:1979ctt}. In particular, the initial data is assumed to be conformally flat and maximally sliced, i.e. $K = 0$. 
However, as also discussed in \cite{Sperhake:2008ga, Shibata:2008rq}, the two boosted black holes are not described by an initially conformally flat hypersurface, leading to spurious gravitational radiation known as junk radiation. 
In \cite{Damour:2014afa, Rettegno:2023ghr}, it was found that the fraction of $E_\mathrm{ADM}$ (and $J_\mathrm{ADM}$) emitted as junk radiation was on the order of $\sim \mathcal{O}(10^{-5})$. 
In Fig.~\ref{fig:weob_junk_comparison}, we demonstrate the effect of junk radiation on our $w^\mathrm{eob}$ scattering angle predictions at 3PM and 4PM. 
At the 3PM and 4PM level, we see that fractional energy differences of $\mathcal{O}(10^{-4})$ and $\mathcal{O}(10^{-3})$ do not have any significant effect. 
It is only at $\mathcal{O}(10^{-2})$ where the effects of junk radiation become important and subsequent analysis would be affected.

\section{NR Simulation Data} \label{Appendix: NR Data}

Here we present the data from our NR simulations, as well as the data from \cite{Damour:2014afa, Rettegno:2023ghr} which has been re-analysed with our new robust scattering angle extraction scheme. Entries of $\theta_\mathrm{NR}$ are accompanied with the dissymmetric error bounds as described in \ref{subsec:nr_scatter_angle_extract}, while dots correspond to a plunge orbit.

Data points accompanied with an asterisk denote scattering angles not used in the analyses of Sec. \ref{Section: Comparisons to NR} (due to unbound/plunge uncertainty), but are still shown in the figures for completeness.

\begin{table}[htbp]
    \caption{Summary of NR data at $P_\mathrm{ADM} = 0.114564$.}
    \setlength{\tabcolsep}{0.4cm}
    \begin{tabular}{c c c c}
    \hline
    \hline   
    {$\Gamma$} & $b$ & $\ell$ & $\theta_{\rm{NR}} \left[\rm deg\right]$ \\
    \hline
    1.02264 & 9.40 & 4.3076 & $\cdots $\\
    1.02264 & 9.50 & 4.3534 & $\phantom{\ast} 368.5819^{+0.0004}_{-11.1042} \ast$\\
    1.02264 & 9.55 & 4.3764 & $328.8896^{+0.0014}_{-2.8654}$\\
    1.02264 & 9.56 & 4.3809 & $322.9123^{+0.0002}_{-2.4425}$\\
    1.02264 & 9.57 & 4.3855 & $317.9898^{+0.0002}_{-2.0382}$\\
    1.02264 & 9.58 & 4.3901 & $313.4235^{+0.0002}_{-1.7243}$\\
    1.02264 & 9.60 & 4.3993 & $305.2217^{+0.0002}_{-1.2791}$\\
    1.02264 & 9.70 & 4.4451 & $274.4404^{+0.0047}_{-0.1506}$\\
    1.02264 & 9.90 & 4.5367 & $236.0760^{+0.4461}_{-0.0313}$\\
    1.02264 & 10.00 & 4.5826 & $222.1698^{+0.6304}_{-0.0008}$\\
    1.02264 & 10.20 & 4.6742 & $201.4541^{+0.0375}_{-0.5979}$ \\
    1.02264 & 10.40 & 4.7659 & $184.9687^{+0.0002}_{-0.5550}$\\
    1.02264 & 11.00 & 5.0408 & $152.0574^{+0.0009}_{-0.5825}$\\
    1.02264 & 12.00 & 5.4991 & $120.7696^{+0.0007}_{-0.2903}$\\
    1.02264 & 13.00 & 5.9573 & $101.6472^{+0.1777}_{-0.0474}$\\
    1.02264 & 14.00 & 6.4156 & $88.3445^{+0.4914}_{-0.0001}$\\
    1.02264 & 15.00 & 6.8739 & $78.4320^{+0.7462}_{-0.0004}$\\
    1.02264 & 16.00 & 7.3321 & $70.6956^{+0.9635}_{-0.0007}$\\
    \hline    
    \hline
    \end{tabular}
    \label{tab:nr_at_padm_0p114564}
\end{table}

\begin{table}[htbp]
    \caption{Summary of NR data at $P_\mathrm{ADM} = 0.15$.}
    \setlength{\tabcolsep}{0.4cm}
    \begin{tabular}{c c c c}
    \hline
    \hline   
    {$\Gamma$} & $b$ & $\ell$ & $\theta_{\rm{NR}} \left[\rm deg\right]$ \\
    \hline
     1.04032 & 7.00 & 4.200 & $\cdots$ \\
     1.04032 & 7.73 & 4.638 & $\phantom{\ast} 390.6213^{+0.0158}_{-6.6477} \ast$ \\
     1.04032 & 7.77 & 4.662 & $339.0783^{+0.0126}_{-1.1475}$\\
     1.04032 & 7.80 & 4.680 & $317.7260^{+0.0111}_{-0.6734}$\\
     1.04032 & 7.87 & 4.722 & $283.7295^{+0.0000}_{-0.1092}$\\
     1.04032 & 7.93 & 4.758 & $263.3659^{+0.2102}_{-0.0245}$ \\
     1.04032 & 8.00 & 4.800 & $244.6663^{+0.8645}_{-0.0001}$ \\
     1.04033 & 8.40 & 5.040 & $184.5080^{+0.1123}_{-0.0001}$ \\
     1.04033 & 8.80 & 5.280 & $153.2602^{+0.0122}_{-0.4977}$\\
     1.04033 & 9.00 & 5.400 & $142.1460^{+0.0125}_{-0.5156}$\\
     1.04033 & 9.40 & 5.640 & $124.8774^{+0.0131}_{-0.3186}$\\
     1.04033 & 9.50 & 5.700 & $121.3506^{+0.0129}_{-0.2765}$\\
     1.04033 & 10.00 & 6.000 & $106.6626^{+0.0249}_{-0.0745}$\\
     1.04033 & 12.00 & 7.200 & $73.4970^{+1.1025}_{-0.0002}$ \\
     1.04033 & 14.00 & 8.400 & $56.8301^{+1.2104}_{-0.0002}$\\
     1.04033 & 16.00 & 9.600 & $46.5893^{+1.2514}_{-0.0003}$\\
    \hline    
    \hline
    \end{tabular}
    \label{tab:nr_at_padm_0p15}
\end{table}

\begin{table}[htbp]
    \caption{Summary of NR data at $P_\mathrm{ADM} = 0.175$.}
    \setlength{\tabcolsep}{0.4cm}
    \begin{tabular}{c c c c}
    \hline
    \hline   
    {$\Gamma$} & $b$ & $\ell$ & $\theta_{\rm{NR}} \left[\rm deg\right]$ \\
    \hline
     1.05548 & 6.93 & 4.851 & $\cdots$\\
     1.05548 & 6.97 & 4.879 & $405.9266^{+0.0472}_{-5.7513} \ast$ \\
     1.05548 & 6.98 & 4.886 & $382.6512^{+0.0441}_{-2.1253}$\\
     1.05548 & 7.00 & 4.900 & $354.3946^{+0.0457}_{-0.9262}$\\
     1.05548 & 7.10 & 4.970 & $285.4616^{+0.04230}_{-0.0570}$\\
     1.05548 & 7.15 & 5.005 & $265.3927^{+0.4638}_{-0.0004}$ \\
     1.05548 & 7.20 & 5.040 & $249.3543^{+0.8857}_{-0.0004}$\\
     1.05548 & 7.40 & 5.180 & $206.7481^{+0.8274}_{-0.0004}$\\
     1.05548 & 7.60 & 5.320 & $180.2335^{+0.0717}_{-0.0004}$\\
     1.05548 & 8.00 & 5.600 & $146.7922^{+0.0519}_{-0.5248}$\\
     1.05548 & 8.50 & 5.950 & $121.4351^{+0.0565}_{-0.2718}$\\
     1.05548 & 9.00 & 6.300 & $104.4585^{+0.1413}_{-0.0233}$\\
     1.05548 & 10.00 & 7.000 & $82.5885^{+0.8514}_{-0.0005}$\\
     1.05548 & 11.00 & 7.700 & $68.7555^{+1.3353}_{-0.0006}$\\
    \hline    
    \hline
    \end{tabular}
    \label{tab:nr_at_padm_0p175}
\end{table}

\begin{table}[htbp]
    \caption{Summary of NR data at $P_\mathrm{ADM} = 0.2$.}
    \setlength{\tabcolsep}{0.4cm}
    \begin{tabular}{c c c c}
    \hline
    \hline   
    {$\Gamma$} & $b$ & $\ell$ & $\theta_{\rm{NR}} \left[\rm deg\right]$ \\
    \hline
     1.07277 & 6.40 & 5.120 & $\cdots$\\
     1.07277 & 6.44 & 5.152 & $405.0183^{+0.1047}_{-1.9022}$ \\
     1.07277 & 6.46 & 5.168 & $367.0004^{+0.1403}_{-0.3949}$\\
     1.07277 & 6.48 & 5.184 & $343.0500^{+0.1053}_{-0.4145}$\\
     1.07277 & 6.50 & 5.200 & $324.8823^{+0.1063}_{-0.4071}$\\
     1.07277 & 6.60 & 5.280 & $269.1780^{+0.6603}_{-0.0006}$\\
     1.07277 & 6.70 & 5.360 & $236.3127^{+1.3311}_{-0.0009}$\\
     1.07727 & 6.80 & 5.440 & $214.3263^{+0.8919}_{-0.0007}$\\
     1.07277 & 7.20 & 5.760 & $160.9356^{+0.0897}_{-0.2894}$\\
     1.07277 & 8.00 & 6.400 & $113.9753^{+0.1017}_{-0.1539}$\\
     1.07277 & 9.00 & 7.200 & $85.6440^{+0.9986}_{-0.0009}$\\
     1.07277 & 10.00 & 8.000 & $69.2309^{+1.5980}_{-0.0010}$\\
     1.07277 & 12.00 & 9.600 & $50.7872^{+2.3631}_{-0.0012}$ \\
     1.07277 & 14.00 & 11.200 & $40.5062^{+2.0473}_{-0.0014}$ \\
     1.07277 & 16.00 & 12.800 & $33.2729^{+1.9880}_{-0.0018}$\\
     1.07278 & 18.00 & 14.400 & $28.3627^{+2.2190}_{-0.0036}$\\
    \hline    
    \hline
    \end{tabular}
    \label{tab:nr_at_padm_0p2}
\end{table}

\begin{table}[htbp]
    \caption{Summary of NR data at $P_\mathrm{ADM} = 0.25$.}
    \setlength{\tabcolsep}{0.4cm}
    \begin{tabular}{c c c c}
    \hline
    \hline   
    {$\Gamma$} & $b$ & $\ell$ & $\theta_{\rm{NR}} \left[\rm deg\right]$ \\
    \hline
    1.11346 & 5.75 & 5.75 & $\cdots$ \\
    1.11346 & 5.76 & 5.76 & $444.8386^{+0.1912}_{-3.8545} \ast$ \\
    1.11346 & 5.77 & 5.77 & $408.3460^{+0.4718}_{-0.0014}$\\
    1.11346 & 5.78 & 5.78 & $387.0918^{+0.6855}_{-0.0014}$\\
    1.11346 & 5.79 & 5.79 & $370.6769^{+0.4963}_{-0.0014}$\\
    1.11346 & 5.80 & 5.80 & $357.1789^{+0.2141}_{-0.0014}$\\
    1.11346 & 5.90 & 5.90 & $281.2531^{+0.5518}_{-0.0014}$\\
    1.11346 & 6.00 & 6.00 & $242.0627^{+1.8561}_{-0.0015}$\\
    1.11346 & 6.30 & 6.30 & $181.1425^{+0.3650}_{-0.0015}$\\
    1.11346 & 6.60 & 6.60 & $149.0268^{+0.2190}_{-0.5148}$\\
    1.11346 & 7.00 & 7.00 & $122.3795^{+0.2298}_{-0.2381}$\\
    1.11346 & 8.00 & 8.00 & $86.5021^{+1.3153}_{-0.0019}$\\
    1.11346 & 9.00 & 9.00 & $67.5560^{+2.3510}_{-0.0022}$\\
    1.11346 & 10.00 & 10.00 & $55.6390^{+2.5752}_{-0.0024}$ \\
    \hline    
    \hline
    \end{tabular}
    \label{tab:nr_at_padm_0p25}
\end{table}

\begin{table}[htbp]
    \caption{Summary of NR data at $P_\mathrm{ADM} = 0.3$.}
    \setlength{\tabcolsep}{0.4cm}
    \begin{tabular}{c c c c}
    \hline
    \hline   
    {$\Gamma$} & $b$ & $\ell$ & $\theta_{\rm{NR}} \left[\rm deg\right]$ \\
    \hline
    1.16174 & 5.3 & 6.36 & $\cdots$  \\
    1.16174 & 5.4 & 6.48 & $399.8354^{+1.2490}_{-0.0014}$ \\
    1.16174 & 5.5 & 6.60 & $299.2155^{+0.2091}_{-0.0311}$\\
    1.16174 & 5.6 & 6.72 & $253.1607^{+1.4554}_{-0.0014}$\\
    1.16174 & 5.8 & 6.96 & $201.6697^{+1.6690}_{-0.0015}$\\
    1.16174 & 6.0 & 7.20 & $171.3999^{+0.2307}_{-0.0015}$ \\
    1.16174 & 6.2 & 7.44 & $150.3628^{+0.2400}_{-0.5347}$\\
    1.16174 & 6.4 & 7.68 & $134.5625^{+0.2444}_{-0.4704}$\\
    1.16174 & 6.6 & 7.92 & $122.1070^{+0.2531}_{-0.1809}$ \\
    1.16174 & 6.8 & 8.16 & $111.9804^{+0.2621}_{-0.0017}$\\
    1.16174 & 7.0 & 8.40 & $103.5455^{+0.4406}_{-0.0018}$\\
    1.16174 & 7.5 & 9.00 & $87.4237^{+1.1162}_{-0.0019}$ \\
    1.16174 & 8.0 & 9.60 & $75.8051^{+1.7169}_{-0.0020}$\\
    1.16174 & 9.0 & 10.80 & $59.9778^{+2.7178}_{-0.0023}$\\
    1.16174 & 10.0 & 12.00 & $49.6201^{+3.3270}_{-0.0026}$\\
    1.16174 & 11.0 & 13.20 & $43.5746^{+3.6969}_{-0.0029}$\\
    \hline    
    \hline
    \end{tabular}
    \label{tab:nr_at_padm_0p3}
\end{table}

\begin{table}[htbp]
    \caption{Summary of NR data at $P_\mathrm{ADM} = 0.35$.}
    \setlength{\tabcolsep}{0.4cm}
    \begin{tabular}{c c c c}
    \hline
    \hline   
    {$\Gamma$} & $b$ & $\ell$ & $\theta_{\rm{NR}} \left[\rm deg\right]$ \\
    \hline
    1.21688 & 5.14 & 7.196 & $\cdots$ \\
    1.21688 & 5.16 & 7.224 & $515.8703^{+0.2086}_{-15.9162} \ast$ \\
    1.21688 & 5.18 & 7.252 & $424.9791^{+1.3519}_{-0.0010}$ \\
    1.21688 & 5.20 & 7.280 & $389.0400^{+1.0642}_{-0.0010}$ \\
    1.21688 & 5.30 & 7.420 & $300.7934^{+0.0362}_{-0.2216}$ \\
    1.21688 & 5.40 & 7.560 & $256.0865^{+2.1209}_{-0.0011}$ \\
    1.21688 & 5.50 & 7.700 & $225.9350^{+2.8570}_{-0.0031}$ \\
    1.21688 & 5.60 & 7.840 & $205.4942^{+1.1508}_{-0.0011}$ \\
    1.21688 & 5.70 & 7.980 & $187.5757^{+0.6256}_{-0.0011}$ \\
    1.21688 & 5.80 & 8.120 & $172.7605^{+0.7473}_{-0.0012}$ \\
    1.21688 & 5.90 & 8.260 & $160.8879^{+0.4679}_{-0.0012}$ \\
    1.21688 & 6.00 & 8.400 & $150.7359^{+0.4576}_{-0.0867}$ \\
    1.21688 & 6.30 & 8.820 & $127.9989^{+0.4621}_{-0.1022}$ \\
    1.21688 & 6.60 & 9.240 & $111.8688^{+0.8739}_{-0.0013}$ \\
    1.21688 & 7.00 & 9.800 & $96.0657^{+1.6432}_{-0.0014}$ \\
    1.21688 & 8.00 & 11.200 & $71.3723^{+3.2471}_{-0.0017}$ \\
    1.21688 & 9.00 & 12.600 & $56.6753^{+4.1586}_{-0.0027}$ \\
    1.21688 & 10.00 & 14.00 & $46.9057^{+4.7019}_{-0.0042}$\\
    
    \hline    
    \hline
    \end{tabular}
    \label{tab:nr_at_padm_0p35}
\end{table}

\section{Relation Between PM Coefficients and $\mathcal{L}$-Resummed/$w$-Potential Coefficients}
\label{app:pm_to_resummed_map}
Following \cite{Damour:2022ybd}, we list the relation between the PM coefficients and the $\mathcal{L}$-resummed coefficients up to 6PM
\begin{align}
    \hat{\theta}_1(\gamma) &= \theta_1(\gamma), \\
    \hat{\theta}_2(\gamma) &= \theta_2(\gamma) - \frac{\ell_0}{2}\theta_1(\gamma), \\
    \hat{\theta}_3(\gamma) &= \theta_3(\gamma) - \frac{\ell_0}{2}\theta_2(\gamma) - \frac{\ell_0^2}{12}\theta_1(\gamma), \\
    \hat{\theta}_4(\gamma) &= \theta_4(\gamma) - \frac{\ell_0}{2}\theta_3(\gamma) - \frac{\ell_0^2}{12}\theta_2(\gamma) - \frac{\ell_0^3}{24}\theta_1(\gamma), \\
    \hat{\theta}_5(\gamma) &= \theta_5(\gamma) - \frac{\ell_0}{2}\theta_4(\gamma) - \frac{\ell_0^2}{12}\theta_3(\gamma) - \frac{\ell_0^3}{24}\theta_2(\gamma) \\ \nonumber &\; \qquad - \frac{19\ell_0^4}{720}\theta_1(\gamma), \\
    \hat{\theta}_6(\gamma) &= \theta_6(\gamma) - \frac{\ell_0}{2}\theta_5(\gamma) - \frac{\ell_0^2}{12}\theta_4(\gamma) - \frac{\ell_0^3}{24}\theta_3(\gamma) \\ \nonumber &\; \qquad - \frac{19\ell_0^4}{720}\theta_2(\gamma) - \frac{3\ell_0^5}{160}\theta_1(\gamma) . \nonumber
\end{align}

The PM coefficients $\theta_k (\gamma)$ can be related to the radial potentials $w_i (\gamma)$ following the iterative procedure in \cite{Damour:2019lcq}, see also \cite{Bern:2019crd,Kalin:2019rwq}. We provide the terms up to 6PM.  
\begin{align}
    \theta_1 (\gamma) &= \frac{1}{2} \frac{w_1 (\gamma)}{\pinf}, \\
    \theta_2 (\gamma) &= \frac{\pi}{4} w_2 (\gamma), \\
    \theta_3 (\gamma) &= -\frac{1}{24}\frac{w^3_3 (\gamma)}{\pinf^3} + \frac{1}{2} \frac{w_1 (\gamma) w_2 (\gamma)}{\pinf} + \pinf w_3 (\gamma), \\
    \theta_4 (\gamma) &= \frac{3 \pi}{8} \left[\frac{1}{2} w^2_2 (\gamma) + w_1 (\gamma) w_3 (\gamma) + \pinf^2 w_4 (\gamma) \right], \\
    \theta_5 (\gamma) &= \frac{1}{160}\frac{w^5_1 (\gamma)}{ \pinf^5} - \frac{1}{12} \frac{w_1^3 (\gamma) w_2 (\gamma)}{\pinf^3} \\ \nonumber &\; + \frac{1}{2} \frac{w_1 (\gamma) w_2^2 (\gamma) + w_1^2 (\gamma) w_2 (\gamma)}{\pinf} \\ \nonumber &\; + 
    2 \pinf \left[ w_2 (\gamma) w_3 (\gamma) + w_1 (\gamma) w_4 (\gamma) \right] \\ \nonumber &\; + \frac{4}{3} \pinf^3 w_5 (\gamma)
    ,\\
    \theta_6 (\gamma) &= \frac{15 \pi}{32} \Bigg[ 
    \frac{1}{3} w_2^2 (\gamma) + 2 w_1 (\gamma) w_2 (\gamma) w_3 (\gamma) \\ \nonumber &\; + w_1^2 (\gamma) w_4 (\gamma) \\ \nonumber &\; + 
    \pinf^2 \left( w_3^2 (\gamma) + 2 w_2 (\gamma) w_4 (\gamma) + 2 w_1 (\gamma) w_5 (\gamma) \right) \\ \nonumber &\; + \pinf^4 w_6 (\gamma)
    \Bigg].
\end{align}

\clearpage

\begin{widetext}

\section{Comparison to NR: Additional Plots}

\subsection{$\mathcal{L}$-Resummation} \label{Appendix: Additional Lresum_plots}

\begin{figure*}[htp]
  \centering
  \subfloat{
      \includegraphics[width=.46\textwidth]{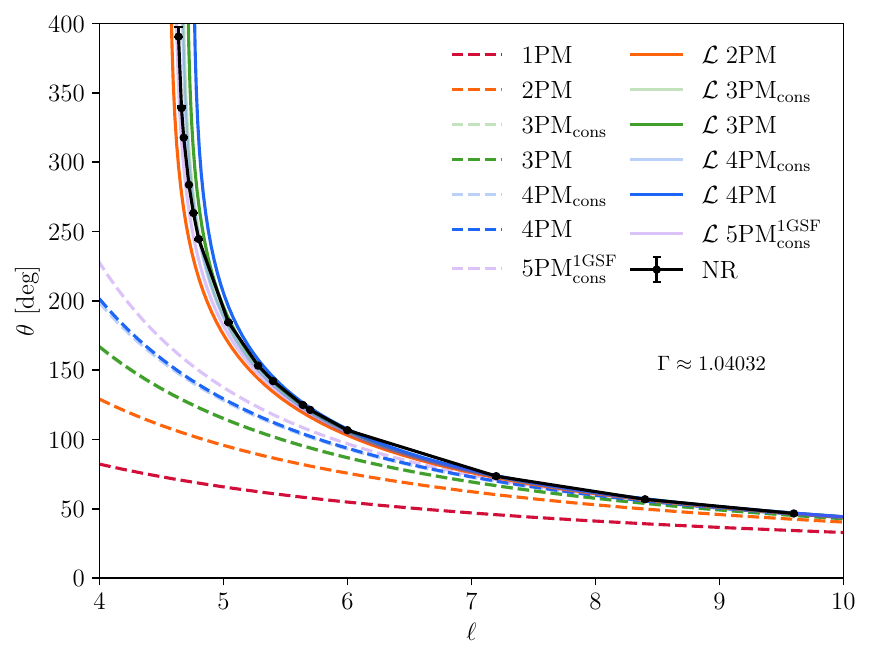}
  }
  \subfloat{
      \includegraphics[width=.46\textwidth]{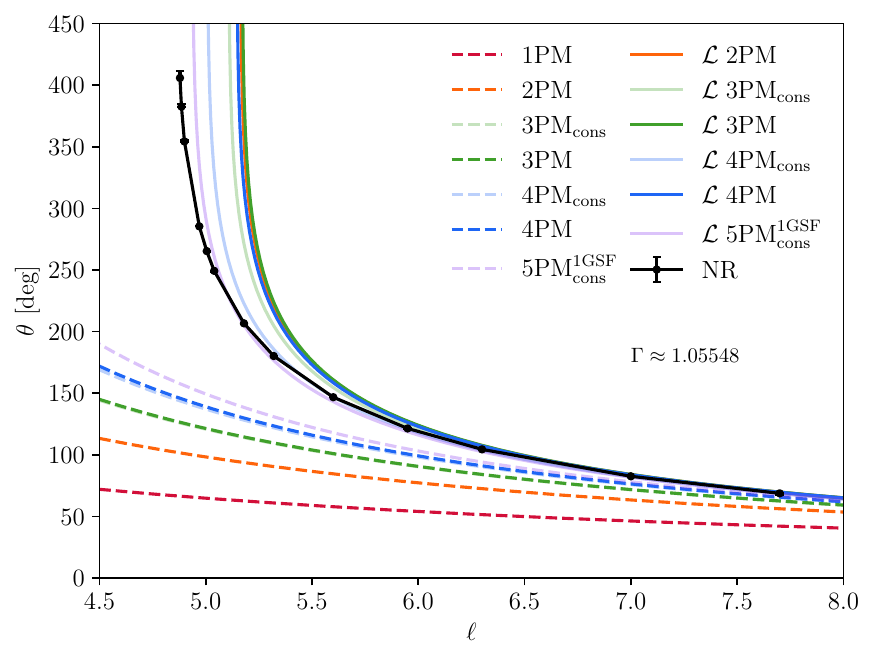}
  }\\   
  \subfloat{
      \includegraphics[width=.46\textwidth]{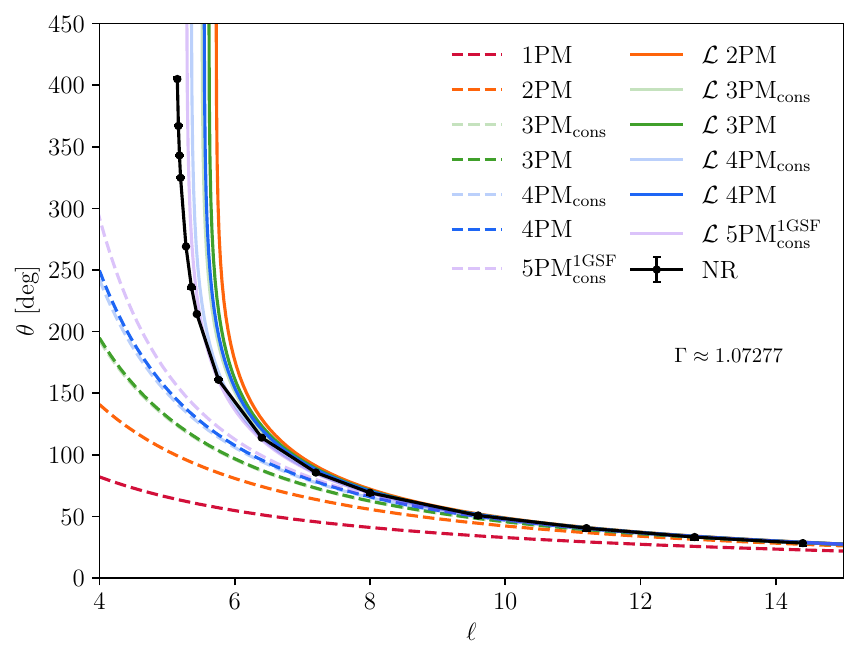}
  }
  \subfloat{
      \includegraphics[width=.46\textwidth]{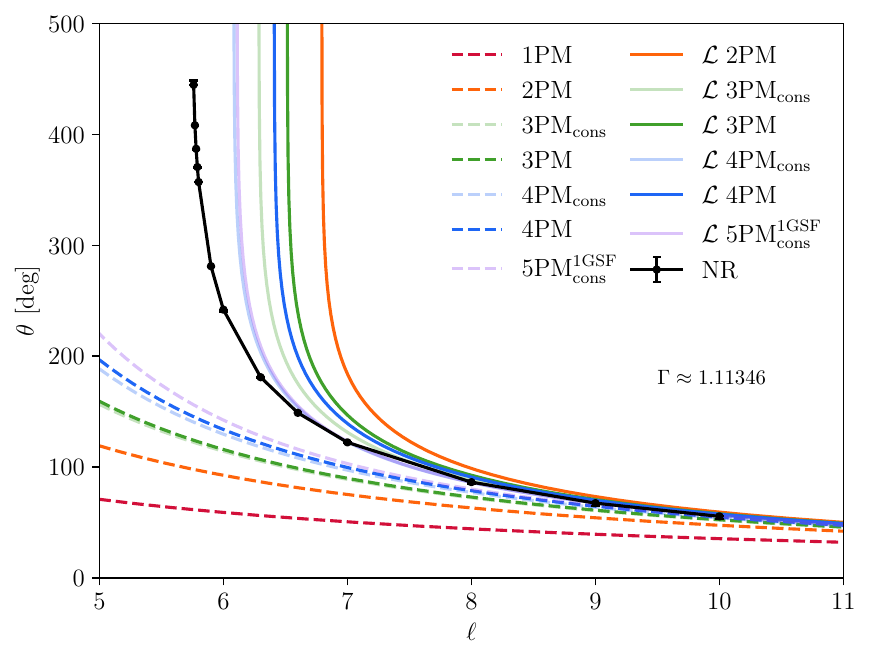}
  }\\
  \subfloat{
      \includegraphics[width=.46\textwidth]{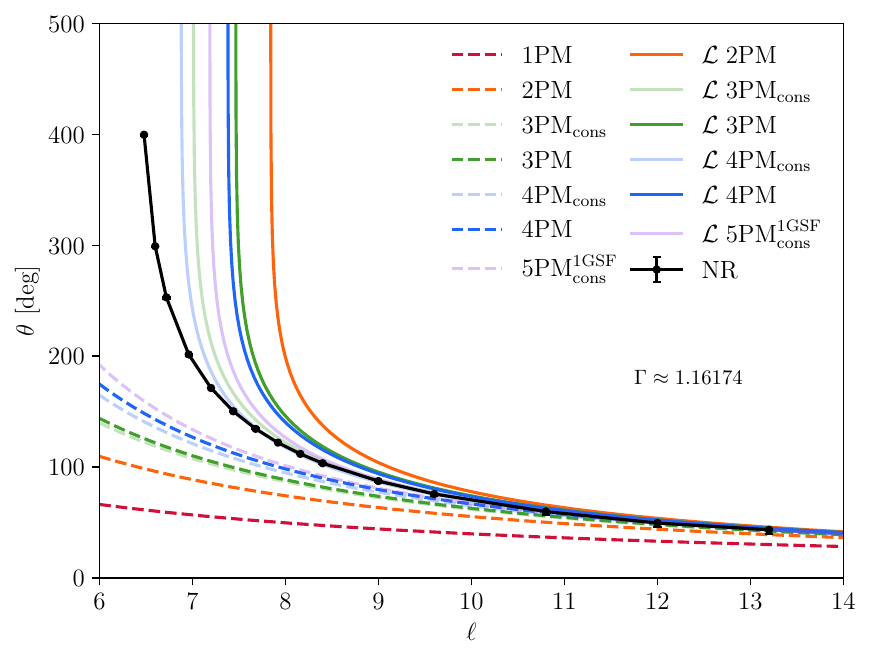}
  }
  \caption{Comparison of $\mathcal{L}$-resummed scattering angle predictions with NR data at energies $\Gamma_2$-$\Gamma_6$. }
  \label{Fig: Additional Lresum_plots}
\end{figure*}

\clearpage

\subsection{The $w^\mathrm{eob}$ Model} \label{Appendix: Additional weob_plots}

\begin{figure*}[htp]
  \centering
  \subfloat{
      \includegraphics[width=.46\textwidth]{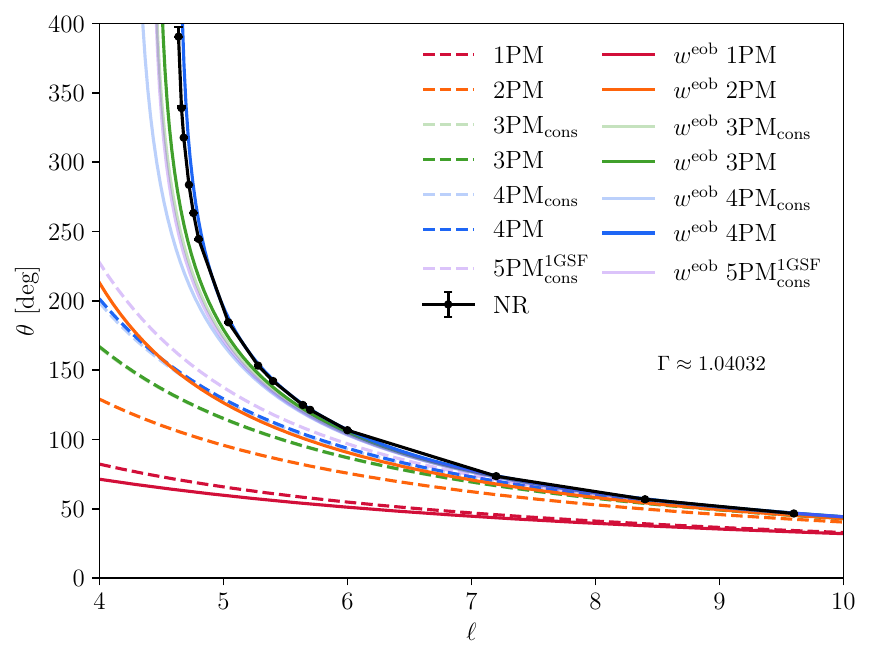}
  }
  \subfloat{
      \includegraphics[width=.46\textwidth]{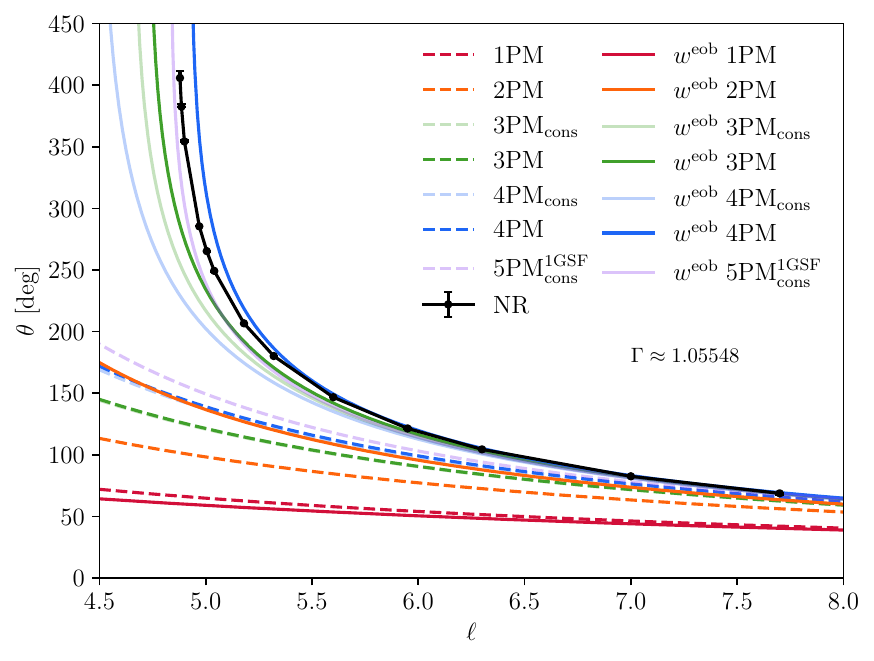}
  }\\ 
  \subfloat{
      \includegraphics[width=.46\textwidth]{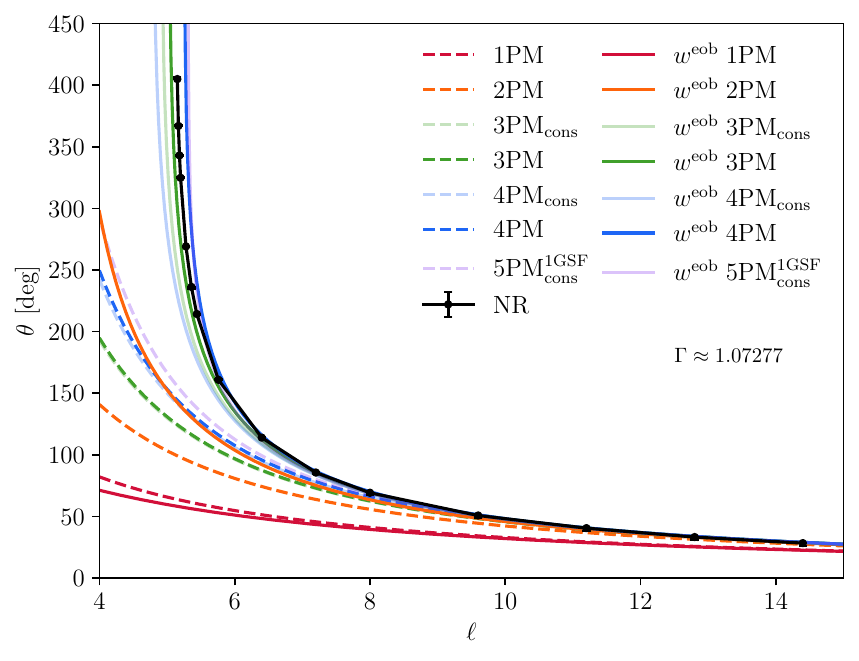}
  }
  \subfloat{
      \includegraphics[width=.46\textwidth]{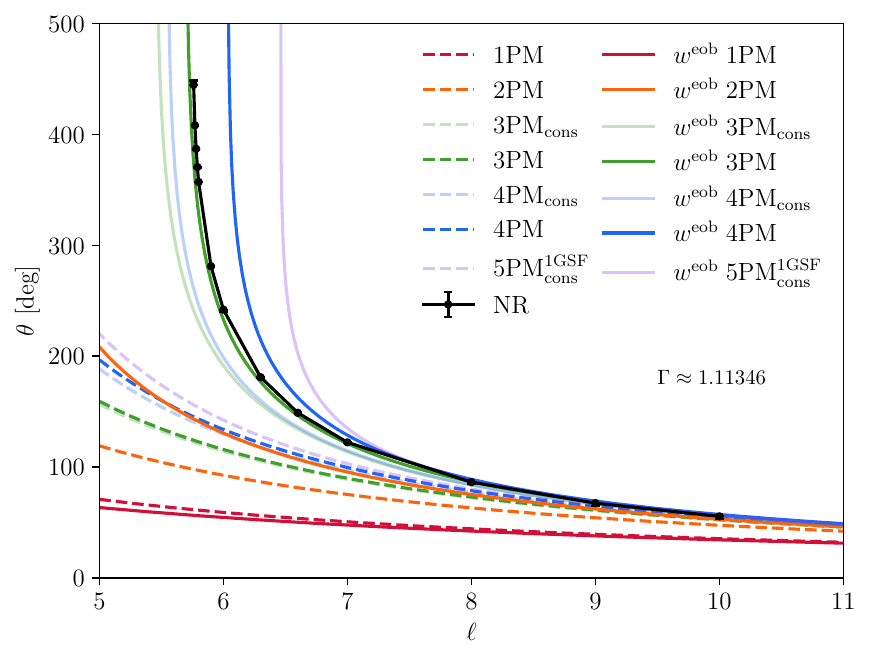}
  }\\
  \subfloat{
      \includegraphics[width=.46\textwidth]{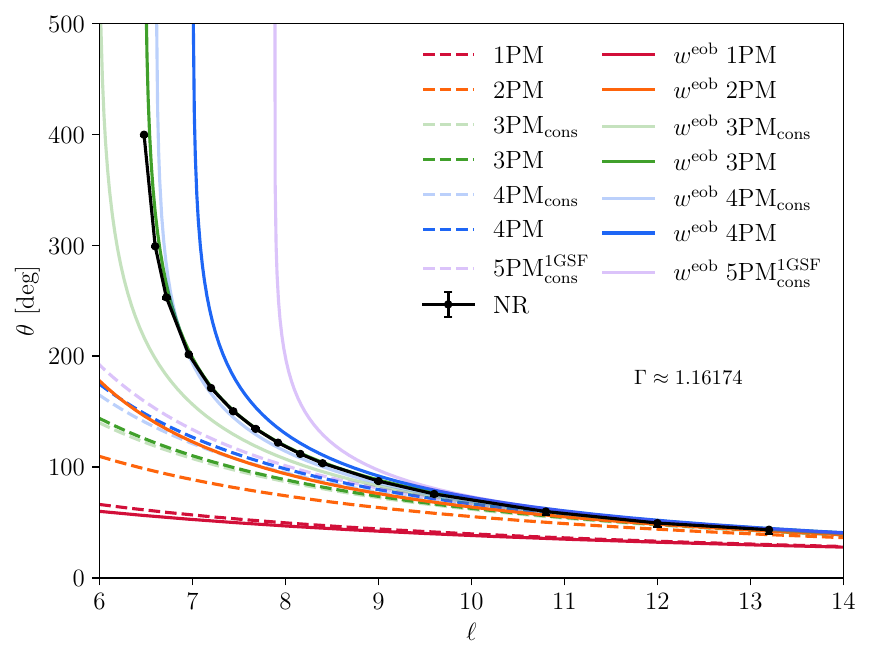}
  }
  \caption{Comparison of $w^\mathrm{eob}$-resummed scattering angle predictions with NR data at energies $\Gamma_2$-$\Gamma_6$.}
  \label{Fig: Additional weob_plots}
\end{figure*}

\clearpage

\subsection{The SEOB-PM Model} \label{Appendix: Additional seob_plots}

\begin{figure*}[htp]
  \centering
  \subfloat{
      \includegraphics[width=.46\textwidth]{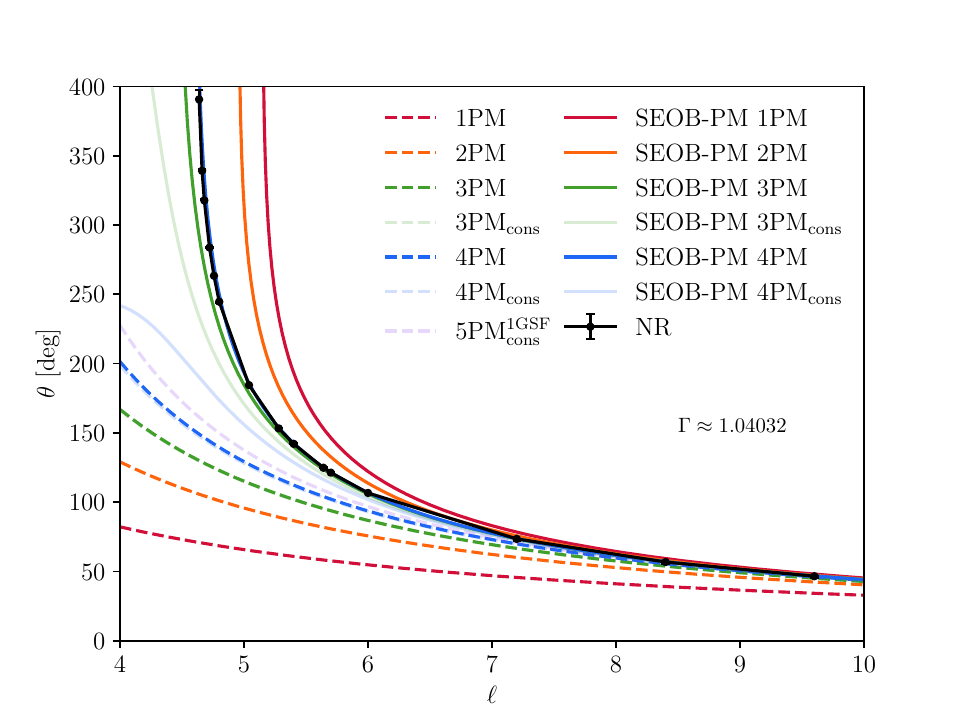}
  }
  \subfloat{
      \includegraphics[width=.46\textwidth]{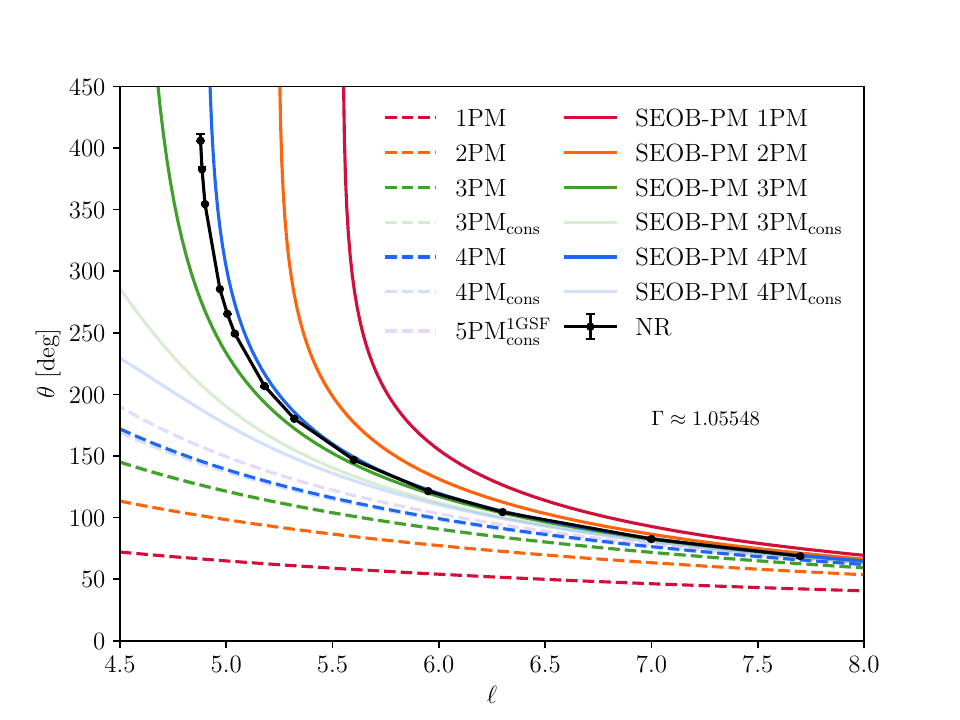}
  }\\ 
  \subfloat{
      \includegraphics[width=.46\textwidth]{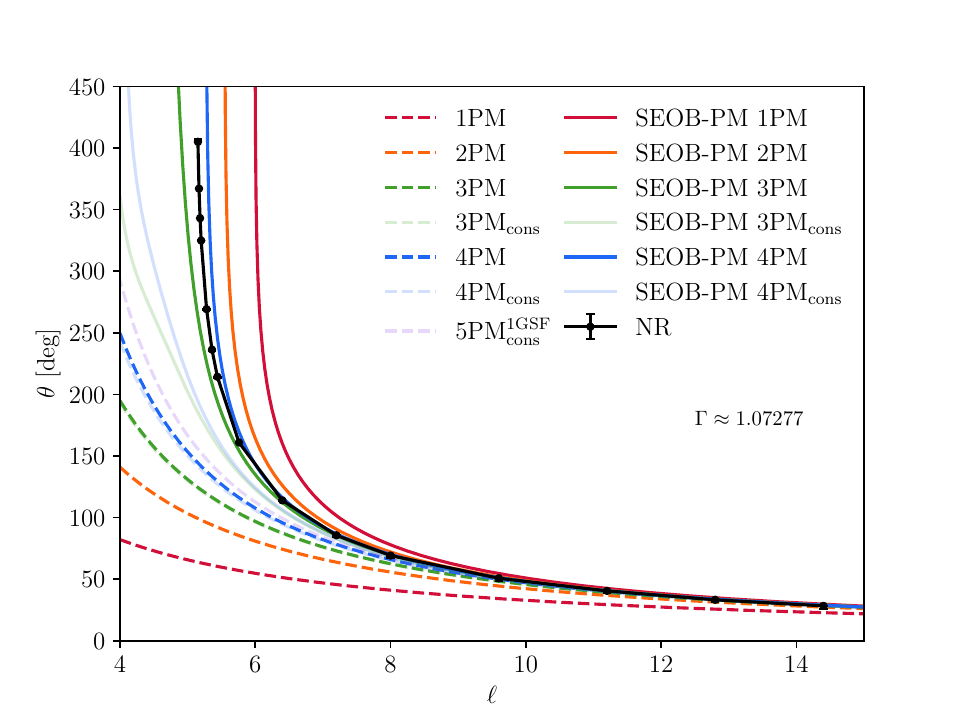}
  }
  \subfloat{
      \includegraphics[width=.46\textwidth]{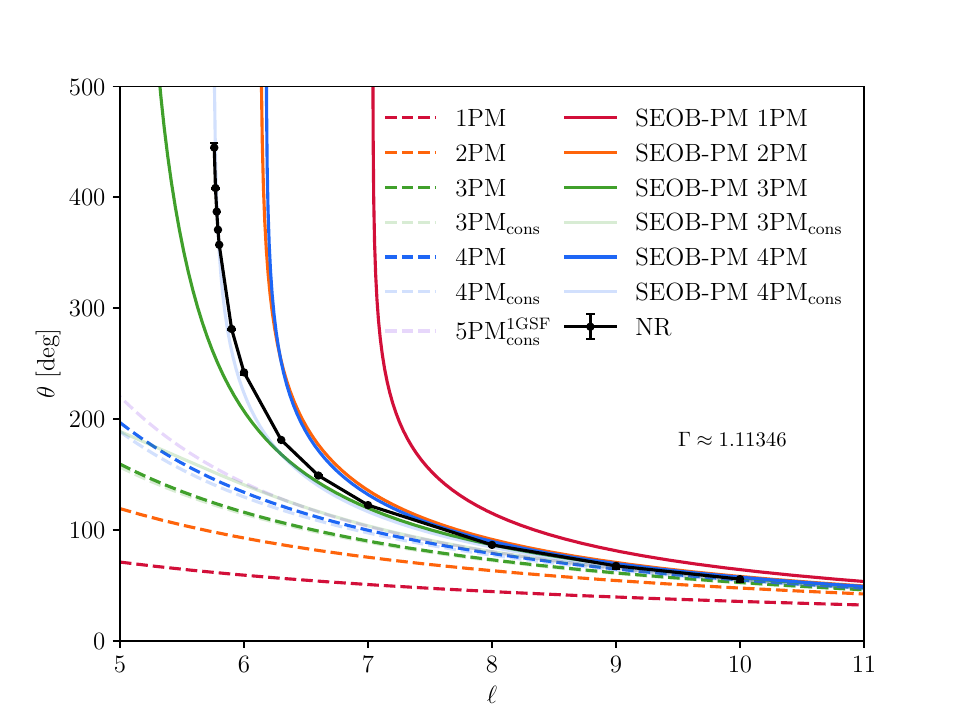}
  }\\
  \subfloat{
      \includegraphics[width=.46\textwidth]{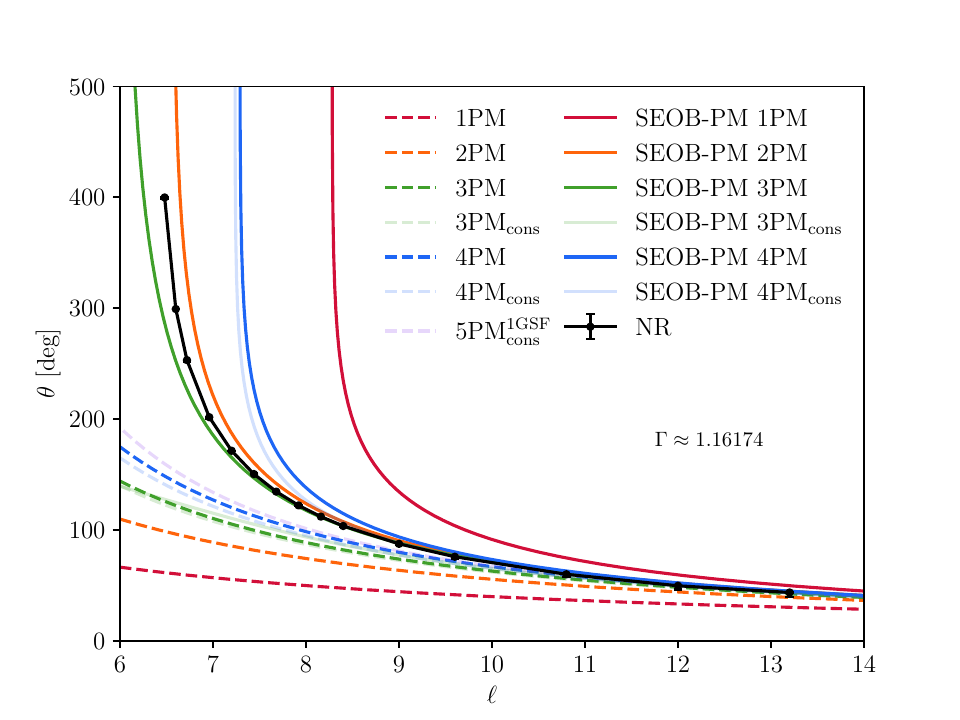}
  }
  \caption{Comparison of SEOB-PM scattering angle predictions with NR data at energies $\Gamma_2$-$\Gamma_6$. 
  }
  \label{Fig: Additional seob_plots}
\end{figure*}

\clearpage

\end{widetext}

\clearpage
\bibliography{references.bib}

\end{document}